\documentclass[fleqn,usenatbib]{mnras}

\usepackage{newtxtext,newtxmath}

\usepackage[T1]{fontenc}

\DeclareRobustCommand{\VAN}[3]{#2}
\let\VANthebibliography\thebibliography
\def\thebibliography{\DeclareRobustCommand{\VAN}[3]{##3}\VANthebibliography}

\usepackage{graphicx}	
\usepackage{amsmath}	
\usepackage{bm}
\usepackage[version=4]{mhchem}
\usepackage{xcolor}
\newcommand{\angstrom}{\textup{\AA}}

\title[On Magnetic Drag]{Geometric Considerations in Hot Jupiter Magnetic Drag Models}

\author[D. A. Christie et al.]{
D. A. Christie,$^{1}$\thanks{E-mail: christie@mpia.de}
T. M. Evans-Soma,$^{2,1}$
N. J. Mayne,$^{3}$
and K. Kohary$^{3}$
\\
$^{1}$Max Planck Institute for Astronomy, K\"onigstuhl 17, D-69117 Heidelberg, Germany\\
$^{2}$School of Information and Physical Sciences, University of Newcastle, Callaghan NSW, 2308, Australia\\
$^{3}$Department of Physics and Astronomy, Faculty of Environment, Science and Economy, University of Exeter, Exeter EX4 4QL, UK
}

\date{Accepted XXX. Received YYY; in original form ZZZ}

\pubyear{2015}

\begin{document}
\label{firstpage}
\pagerange{\pageref{firstpage}--\pageref{lastpage}}
\maketitle

\begin{abstract}
Magnetic fields are expected to impact the atmospheric dynamics of hot and ultra-hot Jupiters due to their increased ionization fractions, compared to that of cooler exoplanets, but our ability to model these magnetic processes is limited by the different coupling regimes between the day and night sides of the planets. One common approach is to approximate the magnetic interactions as a drag acting on the atmosphere.  In this work, we examine, within the context of this drag approximation, the impact of including vertical and meridional drag, in addition to zonal drag, from a background dipole magnetic field on the flows in hot Jupiter atmospheres as well as a relaxation of the assumption of solely meridional currents and demonstrate that the inclusion of meridional and vertical drag can limit flows over the poles in hotter atmospheres, something not seen in models that only consider zonal drag, and the assumption of only meridional currents results in an underestimation of the equatorial drag in all cases examined.
\end{abstract}




\section{Introduction}

Much of our understanding of exoplanets is anchored in what we know about the planets within the solar system; however, we don't always have such analogues available. For example, while we have examples of low density gas shaped by the planetary magnetic field -- the appropriately named magnetospheres -- we do not observe a coupling between the magnetic fields and the weather layers of the solar system planets. Hot Jupiters, close-in, Jupiter-sized planets with temperatures in excess of $1000\,\mathrm{K}$, have the potential for a partial coupling between any planetary magnetic field and the atmospheric gas through collisions between thermally- and photo-ionized atoms in the atmosphere and the bulk neutral gas \citep{perna_2010a,koskinen_2014} that could shape the super-rotating jets expected to dominate their atmospheres.  Whether the magnetic field will meaningfully impact the dynamics of the atmosphere around the photosphere, however, depends on the strength of the planetary magnetic field. Thus far, there have been no direct detections of the magnetic field of a hot Jupiter, and the majority of studies of the weather and climate of these planets have either completely neglected magnetic fields, or used highly simplified approaches. 

It has been proposed that cyclotron emission caused by the interaction between the stellar wind and the planetary field, analogous to auroral emission seen in the solar system, could be observed for hot Jupiters \citep{farrell_1999,zarka_2001,zarka_2007,turner_2019} and thus provide constraint on their magnetic field strengths; however, so far only upper limits have been placed on exoplanetary radio emission \citep[e.g.][]{griessmeier_2017,cendes_2022}. Indirect estimates of magnetic field strengths based on observational data \citep[e.g.,][]{vidotto_2010,cauley_2019} and theoretical estimates based on scaling arguments \citep[e.g.][]{christensen_2009,yadav_2017} yield potential field strengths in excess of $100\,\mathrm{G}$, although direct, independent confirmation remains elusive.  A more recent proposal by \citet{savel_2024} has suggested that observations of the relative motion between ions and neutrals could be used to infer the magnetic field strength.\footnote{We discuss this approach in Appendix \ref{Apdx:DriftSpeed}}  Attempts instead to match observations of hot and ultra-hot Jupiters with synthetic observations from general circulation models (GCMs) have in some cases favoured the introduction of a strong atmospheric drag \citep[e.g.][]{kreidberg_2018b,may_2021b,beltz_2023}, and although some of these models can be agnostic to the underlying mechanism, magnetic effects, approximated in their influence by a drag term, are still a compelling explanation. This approach to indirectly inferring planetary magnetic field strengths requires confidence that our models properly capture the impact of magnetic fields on the atmospheres being investigated, both regarding whether magnetic drag is accurately captured and whether magnetic drag is an appropriate approximation of magnetic effects.


The modelling of magnetic interactions in the weather layers of hot Jupiters specifically\footnote{\citet{rogers_2017} and \citet{rogers_2017b} present simulations that solve the equations of non-ideal magnetohydrodynamics with a horizontally-varying resistivity; however, the temperature regime investigated is that of ultra-hot Jupiters with nightside temperatures in excess of 1800\,K, keeping the nightside temperatures sufficiently ionized to make the problem tractable.} presents a number of challenges due to the disparate conditions found between the day and night sides of these planets. The nightside, with exponentially lower ionization, should be entirely decoupled from the field around the photosphere while the dayside experiences a partial coupling, depending on the field strength, gas temperature, and ionization state.  While each of these regimes can be modelled individually, the modelling of an atmosphere encompassing both regimes simultaneously is challenging. A number of approaches have been taken to simplify the problem and make it tractable.  


One approach is to assume a horizontally-uniform resistivity, artificially increasing the coupling of the nightside to the magnetic field where it would otherwise be decoupled due to the low ionization fraction, allowing for the equations of non-ideal magnetohydrodynamics to be solved \citep[e.g.,][]{rogers_2014,rogers_2014b}.  This approach permits the deformation of the magnetic field by the winds to be followed and the evolution of the field to feedback on the atmosphere (see Figure 1 of \citealt{rogers_2017b} for a representation of the deformed field lines from their simulations).  However, studies of this form have made use of the anelastic \citep{rogers_2014,rogers_2014b,rogers_2017b} or Boussinesq \citep{batygin_2013} approximations to suppress sonic and magnetosonic waves, as well as differing thermal forcing and the inclusion of thermal diffusion, further limiting the formation of equatorial jets and resulting in flows in the hydrodynamic limit different to those seen in general circulation models. For example, the hydrodynamic (i.e., non-magnetic) GCM simulations of HD\,209458\,b in \citet{showman_2008} have peak mean zonal windspeeds in excess of $4\,\mathrm{km\,s^{-1}}$ compared to $\sim 0.5\,\mathrm{km\,s^{-1}}$ in the hydrodynamic models of \citet{rogers_2014}. Thus any comparison of the impact of magnetic fields between these models and GCMs is subject to the caveat that the underlying hydrodynamic models do not agree.   A different method of modelling the atmosphere while still solving the magnetohydrodynamic equations was taken by \citet{hindle_2019,hindle_2021} wherein the induction equation is solved in a shallow atmosphere limit, effectively modelling a single scale height and assuming the vertical field component  of the magnetic field is small relative to the azimuthal component. They find magnetically-induced wind reversals (i.e., westward winds in the magnetic case compared to eastward winds in the non-magnetic case), and argue it may be responsible for the proposed variable hot spot shift on HAT-P-7\,b \citep{armstrong_2016}, although the variability may be a signal from supergranulation on the surface of the host star or another astrophysical or systematic source of noise  unrelated to variability in the planetary atmosphere \citep[][]{lally_2022}.  All of these approaches provide insight into the magnetic field geometry on the dayside as the flow drags it around the planet at the expense of not capturing the decoupling on the nightside of the planet due to the assumption of horizontally-uniform resistivity.  

A complementary approach, which is adopted in this study, is to assume a static background field, usually a dipole, abandoning any attempts to solve the equations of non-ideal magnetohydrodynamics or otherwise follow the evolution of the magnetic field, instead reducing the magnetic interaction to a linear drag dependent on the local temperature and pressure included when solving the hydrodynamic equations\footnote{In models of gaseous exoplanets, horizontally-isotropic drag is often introduced throughout the computational domain or within a specific region (e.g., deep in the atmosphere) to maintain numerical stability or match observations \citep[e.g.,][]{kreidberg_2018b,tan_2019}.  While magnetic fields are sometimes invoked as a potential source for the drag, no attempt is made within the implementations of the models to capture any aspect unique to magnetic fields, and as such, we do not include them in the discussion of {\em magnetic} drag models.}. This approach was introduced for hot Jupiters in \citet{perna_2010a,perna_2010b} based on the estimates of magnetic drag deep in the atmosphere of Jupiter from \citet{liu_2008}, and has since been used in several studies \citep[e.g.][]{rauscher_2013,beltz_2022,beltz_2023,kennedy_2025,beltz_2025}.  Although not following the evolution of the magnetic field is a limiting assumption that cannot be ignored and approximating the magnetic interactions as a drag term is only strictly applicable in regimes where the magnetic field is weakly coupled to the neutral atmosphere, it does allow for the models to capture both a decoupled night side and partially coupled day side and progress our understanding of the magnetospheric interaction for hot Jupiters. 

\begin{figure}
    \centering
    \includegraphics[width=3.1in]{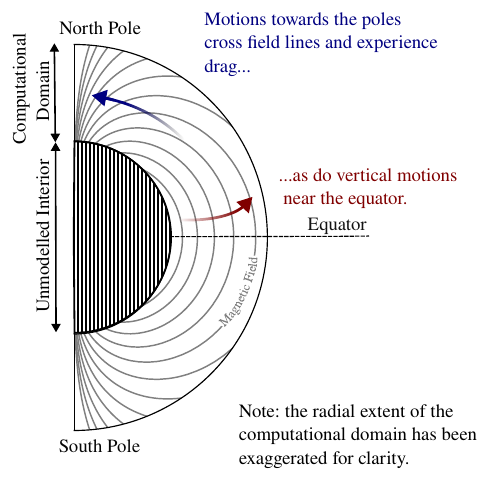}
   \caption{An illustration highlighting how meridional motions away from the equator and vertical motions near the equator experience drag as they cross the field lines of the planetary dipole. Note that in the figure, the radial extent of the computational domain has been exaggerated for clarity.  }
    \label{fig:dipole}
\end{figure}

Beyond the limitations of not following the evolution of the magnetic field, magnetic drag models in hot Jupiter atmospheres, due to their provenance in models of drag in the atmosphere of Jupiter, assume the winds are primarily zonal, and, as such, assume that the drag similarly can be assumed to be zonal.  As the ionization of the atmosphere increases and with it, the coupling between the magnetic field and atmosphere, the zonal drag can become sufficiently strong so as to inhibit the formation of an equatorial jet and instead transition the flow from a zonal flow to a meridional flow over the poles \citep{beltz_2022}. This transition, however, violates the underlying assumption regarding the geometry of the flow that is used to justify the assumption of drag applying only in the zonal direction. This fundamental inconsistency necessitates consideration of the full three dimensional geometry of the magnetic field (see Figure \ref{fig:dipole}). 

In this work, we build on the work of \citet{rauscher_2013} and \citet{beltz_2022} by relaxing the assumption of a zonal flow and solely meridional currents in the calculation of the magnetic drag and to consider the full magnetic field geometry.  We present a model  including vertical and meridional magnetic drag in a hot Jupiter atmosphere and demonstrate that for sufficiently hot atmospheres where the zonal drag can inhibit equatorial jet formation, the meridional and vertical components, combined with the zonal component, shape the flow, creating a diverging day-night flow and potentially creating a hot-spot of nearly static gas around the substellar point. We also demonstrate that in all cases examined, the assumption of solely meridional currents significantly underestimates the equatorial drag. The paper is structured as follows: in Section \ref{Sec:DragModels}, we summarize how previous magnetic drag models have been implemented and outline our approach to extending these models to include the full magnetic field geometry and non-zonal flows. In Section \ref{Sec:Simulations}, we detail how this drag model is implemented in a general circulation model and present the simulation results in \ref{Sec:Results}. We elaborate on the implications of these results and provide a discussion of the limitations of the model in Section \ref{Sec:Conclusions}.  In the Appendix, we provide a derivation of the relative velocity between ions and neutrals within the atmosphere which is notably different to that presented in \citet{savel_2024}.


\section{Magnetic Drag Models}

\label{Sec:DragModels}
In this section, we summarize magnetic drag models as they have been used previously to study magnetic effects in the atmospheres of hot Jupiters and outline how we expand on these models to account for the magnetic field geometry and relax the assumption of solely meridional currents. Motivated by the coordinate choices in the {\sc Unified Model} general circulation model, discussed further in Section \ref{SubSec:UM}, we adopt a spherical coordinate system with the equator corresponding to a latitude of $\theta=0^\circ$ and the poles located at $\theta=\pm 90^\circ$.  The anti-stellar point is at a longitude of $\lambda=0^\circ$ and the substellar point is at $\lambda=180^\circ$.  The radial coordinate is $r$ with the origin at the centre of the planet.

\subsection{Calculating the Ionization and Conductivities}
\label{SubSec:Conductivity}
The magnetic force on the atmosphere is mediated via collisions between the charged and neutral species.  The drag timescales used in the simulations  will depend on the ionization state of the atmosphere through the parallel, perpendicular, and Hall resistivities ($\eta_\parallel$, $\eta_\perp$, and $\eta_\mathrm{H}$) and conductivities ($\sigma_\parallel$, $\sigma_\perp$, and $\sigma_\mathrm{H}$).  The resistivities and conductivities relate the electric field and the current via Ohm's law, 

\begin{equation}
\mathbf{E}_n = \eta_\parallel \mathbf{j}_\parallel + \eta_\perp \mathbf{j}_\perp + \eta_\mathrm{H}\mathbf{j}\times \mathbf{b}\,\, ,
\label{Eqn:OhmsLaw}
\end{equation}
\noindent or its inverse,
\begin{equation}
\mathbf{j} = \sigma_\parallel \mathbf{E}_{\mathrm{n},\parallel} + \sigma_\perp \mathbf{E}_{\mathrm{n},\perp} - \sigma_\mathrm{H}\mathbf{E}_\mathrm{n}\times \mathbf{b}\,\, .
\label{Eqn:j}
\end{equation}

\noindent Here $\mathbf{E}_\mathrm{n}$ is the electric field in the frame of the neutral gas, $\mathbf{j}$ is the electric current and $\mathbf{b}$ is a unit vector in the direction of the local magnetic field, with the electric field $\mathbf{E}_n$ further decomposed into components parallel ($\mathbf{E}_\mathrm{n,\parallel}$) and perpendicular ($\mathbf{E}_\mathrm{n,\perp}$) to the magnetic field, and the current  $\mathbf{j}$ is similarly decomposed into components parallel ($\mathbf{j}_\parallel$) and perpendicular ($\mathbf{j}_\perp$) to the magnetic field.      

We compute the resistivities and conductivities from the ionization state of the atmosphere assuming it has solar metallicity\footnote{Elemental abundances are taken from \citet{asplund_2009}.} and is in thermo-chemical equilibrium,  without photo-ionization.  In this case, the population of ions is dominated by \ce{K+} and \ce{Na+} due to the low ionization potentials and relatively high elemental abundances of \ce{Na} and \ce{K} with the number density of electrons computed from charge neutrality.   The individual fractional ionization of species $i$ ($x_\mathrm{e,i}$) is computed using the Saha equation,

\begin{equation}
\frac{x_{\mathrm{e,i}}^2}{1+x_\mathrm{e,i}^2} = \frac{1}{n_i}\left(\frac{2\pi m_\mathrm{e}k_\mathrm{B}T}{h^2}\right)^{3/2}\exp\left(-\frac{\epsilon_i}{k_\mathrm{B}T}\right)  \,\, ,
\end{equation}

\noindent where $\epsilon_i$ is the ionization potential of the neutral atom, $m_\mathrm{e}$ is the mass of an electron, $k_\mathrm{B}$ is the Boltzmann constant, $T$ is the gas temperature, $h$ is the Planck constant, and $n_i$ is the elemental abundance of species $i$.   The total ionization fraction of the gas is then given by

\begin{equation}
    x_\mathrm{e} = \sum_i \frac{n_i}{n}x_\mathrm{e,i}\,\, ,
\end{equation}

\noindent where $n = P/kT$ is the total number of particles.  This approach, used in \citet{menou_2012b}, treats each species separately and will overestimate the ionization as it doesn't consider the contribution electrons from other species; however, the error is less than a factor of two based on our tests, and is computationally expedient. 

The conductivities are then computed following, e.g., \citet{parks_1991},

\begin{align}
\sigma_s & = \frac{n_s q_s^2\tau_{s\mathrm{n}}}{m_s}\,\, , \\
\sigma_\parallel & = \sum_s \sigma_s\,\, , \\
\sigma_\perp & = \sum_s \frac{\sigma_s}{1+\left(\omega_s\tau_{s\mathrm{n}}\right)^2}\,\, , \\
\sigma_\mathrm{H} & = -\sum_s \frac{\sigma_s\omega_s\tau_{s\mathrm{n}}}{1+\left(\omega_s\tau_{s\mathrm{n}}\right)^2}\,\, , \label{Eqn:sigHall}
\end{align}

\noindent with the corresponding resistivities

\begin{align}
\eta_\parallel & = \sigma_\parallel^{-1} \label{Eqn:etaparl}\\
\eta_\perp & = \frac{\sigma_\perp}{\sigma_\perp^2 + \sigma_\mathrm{H}^2} \label{Eqn:etaperp}\\
\eta_\mathrm{H} &= \frac{\sigma_\mathrm{H}}{\sigma_\perp^2 + \sigma_\mathrm{H}^2}\,\, ,\label{Eqn:etaH}
\end{align}

\noindent where $n_s$ is the abundance of species $s$ with corresponding mass $m_s$ and charge $q_s$, $\omega_s=q_sB/m_sc$ is the gyrofrequency of species $s$, and $\tau_{s\mathrm{n}}$ is the momentum transfer timescale for species $s$ in a gas of neutrals.  The momentum transfer timescale for species $s$ in a gas of neutrals $n$ is 

\begin{equation}
\tau_{sn} = \frac{m_s + m_n}{m_n}\frac{1}{n_n\left<\sigma w\right>_{sn}}\,\,,
\end{equation}

\noindent where $n$ is the neutral species.  As we are considering an atmosphere dominated by \ce{H2} and \ce{He}, the total momentum transfer timescale between species $s$ and the neutral medium is 

\begin{equation}
\frac{1}{\tau_{s\mathrm{n}}}= \frac{1}{\tau_{s\mathrm{H_2}}} + \frac{1}{\tau_{s\mathrm{He}}}.
\end{equation}

The momentum transfer rate coefficient $\left<\sigma w\right>_{sn}$ between the ion species $s$ -- either \ce{K+} or \ce{Na+} for the purposes here -- and the neutral species $n$ -- \ce{H2} or \ce{He} -- quantifies the rate of momentum transfer between two Maxwellian gases with non-zero relative mean velocities and is given by the Langevin approximation 

\begin{equation}
\left<\sigma w\right>_\mathrm{sn} = 2.81\times 10^{-9}\left(\frac{p_n}{\angstrom^3}\right)^{1/2}\left(\frac{m_\mathrm{H}}{m_\mathrm{red}}\right)^{1/2}\,\, \mathrm{cm^3\,s^{-1}},
\end{equation}

\noindent where $p_n$ is the polarization of the neutral species, and $m_\mathrm{red}$ is the reduced mass.  For collisions between electrons and neutral species, we use momentum transfer rate coefficients  from \citet{pinto_2008}.

Previous examinations of magnetic drag \citep[e.g.][]{rauscher_2013,beltz_2022,beltz_2023,kennedy_2025} have assumed the Ohmic limit where the resistivity and conductivity can be treated as scalars ($\eta_\perp \simeq \eta_\parallel \equiv \eta$ and $\eta_\mathrm{H} \ll \eta$ or equivalently $\sigma_\perp \simeq \sigma_\parallel \equiv \sigma$ and $\sigma_\mathrm{H} \ll \sigma$) and used the resistivity from \citet{perna_2010a},\footnote{Due to differing definitions of the resistivity, we have included a factor of $4\pi/c^2$ to make units consistent with the rest of this paper.  We also note that this expression for the resistivity is the same as in \citet{menou_2012b} which is the more frequently cited source.}  

\begin{equation}
\eta_\mathrm{{Perna}} = \sigma_\mathrm{e}^{-1} =  3.271\times 10^{-18}\frac{\sqrt{T/1\,\mathrm{K}}}{x_\mathrm{e}}\,\, \mathrm{s}\,\,,
\label{Eqn:etaPerna}
\end{equation}

\noindent based on the electron conductivity of the gas.   This agrees well with the parallel resistivity calculated in Equation \ref{Eqn:etaparl} and remains applicable so long as the gas is in the Ohmic regime ($P \gtrapprox 10^{-3}\,\mathrm{bar}$ for the conditions investigated in this work).  At lower pressures, the perpendicular conductivity and resistivity diverge from their parallel counterparts, as the gas is no longer in the Ohmic regime.  We show how this results in longer drag timescales in Section \ref{SubSec:Extending}.    It should also be noted that in the Ohmic regime, there is no obvious preference between presenting the drag timescale in terms of the resistivity versus presenting it in terms of conductivity as they are easily interchangeable ($\eta = \sigma^{-1}$).  As a result, we follow the convention adopted in \citet{perna_2010a} and used in subsequent magnetic drag papers \citep[e.g.,][]{menou_2012b,rauscher_2013} to work in terms of the resistivity $\eta$ when discussing zonal drag models in the next section.   When generalising the drag model in Section \ref{SubSec:Extending} and considering conditions outside of the Ohmic regime, it is natural to work in terms of conductivities $\sigma_\parallel$, $\sigma_\perp$, and $\sigma_\mathrm{H}$.

\subsection{The Zonal Drag Model}
Zonal magnetic drag models for hot Jupiters were introduced in \citet{perna_2010a} as a way to capture the impact of magnetic fields in a weakly-coupled limit where directly solving the equations of non-ideal magnetohydrodynamics can become prohibitively computationally expensive. They assume that there exists a static background dipole magnetic field aligned with the rotational axis of the planet that exerts force on the partially-ionized atmosphere of the planet.    The magnetic forces on the atmosphere depend on the local magnetic field $\mathbf{B}$ and local currents $\mathbf{j}$,  
\begin{equation}
    \left(\rho \frac{\partial \mathbf{u}}{\partial t}\right)_\mathrm{magnetic} = \frac{1}{c}\left(\bm{j}\times\bm{B} \right) = \frac{1}{c}\left(\bm{j}_\perp\times\bm{B} \right) \,\, ,
    \label{Eqn:BForce}
\end{equation}
\noindent where $\rho$ is the gas density and $\mathbf{u}=(u,v,w)$ is the velocity with $u$, $v$, $w$ being the individual zonal, meridional, and radial components, respectively. Based on a similar model of magnetic drag on a radially-varying angular velocity $\Omega$  deep in the atmosphere of Jupiter \citep{liu_2008}, it is assumed that the winds are zonal and that the currents are meridional, with $\mathbf{j}\times\mathbf{B}$ becoming $\mathbf{j}_\theta\times\mathbf{B}$.  Based on Equation \ref{Eqn:BForce}, the characteristic stopping timescale for a zonal wind with speed $u$ can be approximated as

\begin{equation}
\tau_\mathrm{drag,Perna} = \frac{\rho|u|c}{|\mathbf{j}_\theta\times\mathbf{B}|}\,\, .
\label{Eqn:PernaStopping}
\end{equation}

The magnitude of the meridional current $|j_\theta|$ can be estimated from the induction equation which governs the evolution of the magnetic field, 
\begin{equation}
    \frac{\partial \bm{B}}{\partial t} = \bm{\nabla}\times \left(\bm{u}\times\bm{B}\right) - c\bm{\nabla}\times\left(\eta_\parallel \mathbf{j}_\parallel + \eta_\perp \mathbf{j}_\perp + \eta_\mathrm{H}\mathbf{j}\times\mathbf{b} ] \right)\,\, .
    \label{Eqn:Induction}
\end{equation}

\noindent It is assumed that the resistivity is Ohmic allowing the last term in Equation \ref{Eqn:Induction} to be simplified, 

\begin{equation}
    \frac{\partial \bm{B}}{\partial t} = \bm{\nabla}\times \left(\bm{u}\times\bm{B}\right) - c\bm{\nabla}\times\left(\eta \mathbf{j}  \right)\,\, .
    \label{Eqn:InductionOhmic}
\end{equation}

\noindent As the magnetic field is assumed to be static ($\partial \mathbf{B}/\partial t = 0$), the magnitude of the meridional current can be approximated from Equation \ref{Eqn:InductionOhmic} as $|j_\theta| \sim |u| B/c\eta$.   This can be combined with Equation \ref{Eqn:PernaStopping}
to arrive at the governing equations of the zonal drag model,

\begin{equation}
\frac{\partial u}{\partial t} = - \frac{u}{\tau_\mathrm{drag,Perna}}\,\, ,
\label{Eqn:Perna}
\end{equation}

\noindent which results in the slowing, and potential suppression, of zonal winds on a timescale of
\begin{equation}
    \tau_\mathrm{drag,Perna} = \frac{c^2\rho\eta}{B^2|\sin\theta|}\,\, 
    \label{Eqn:orig_drag}
\end{equation}

\noindent which due to the temperature dependence of the resistivity by way of the ionization fraction, primarily acts on the dayside of the atmosphere. The factor of $\sin\theta$ in the drag timescale comes from cross product of $\mathbf{j}_\theta$ and $\mathbf{B}$ in Equation \ref{Eqn:PernaStopping} and is a result of the assumption of solely meridional currents, thus any drag on the gas at the equator has to come from radial or zonal currents.  This is exactly where the assumption of meridional currents breaks down in the model of \citet{liu_2008} and \citet{perna_2010a}, as symmetry leads to meridional currents vanishing at the equator as well as at the poles and continuity thus requiring radial currents.  These radial currents then exert a drag on the zonal flow (see Equation \ref{Eqn:BForce})\footnote{Appendix A of \citet{liu_2008} contains a derivation of the currents for a simple toy model. Their Figure 14 illustrates how currents can become radial at both the equator and at the poles. }.


The assumption of solely meridional currents also precludes the possibility of drag on meridional winds and thus does not capture the full magnetic field geometry.   In the case of a dipole, meridional flows near the equator will not experience drag as the motions are parallel to the local magnetic field; however, near the poles, the magnetic field becomes vertical which should result in a suppression of meridional motions (see Figure \ref{fig:dipole}). Moreover, near the equator, the unperturbed magnetic field lines are horizontal, requiring any vertical motions to cross the magnetic field, thus potentially experiencing drag.  While the zonal drag is assumed to be the primary limiter on the equatorial jet speed, this suppression of vertical and meridional motions has the potential to further impact the vertical and meridional eddy momentum fluxes responsible for driving the jet, as outlined in \citet{showman_2011a}. Put simply, previous works assuming a current only in the meridional direction, and consequently, a purely zonal drag,  will break down as the drag becomes significant. 

\subsection{Extending the Drag Model}
\label{SubSec:Extending}
\begin{figure}
    \centering
    \includegraphics{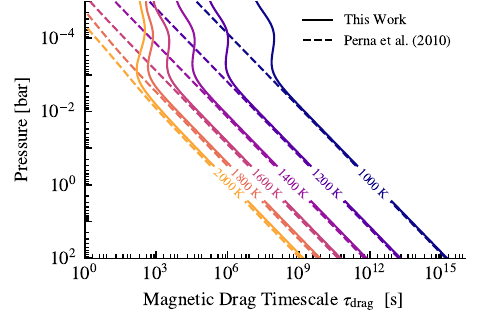}
    \caption{The magnetic drag timescale computed using the perpendicular conductivity (see Equation \ref{eqn:drag3d}) for a $10\,\mathrm{G}$ magnetic field for a number of different temperatures (solid lines).  For comparison, the drag timescales computed from \citet{perna_2010a} using the same ionization fractions (dashed lines). }
    \label{fig:drag_timescales}
\end{figure}

To generalize the magnetic drag model, we alter it in two ways:  first, we relax the assumption that the gas is in the Ohmic regime and compute the conductivities as in Section \ref{SubSec:Conductivity}. Second, we allow for drag in all three directions to better account for the full magnetic field geometry. These changes will generalize the drag formulation (Equation \ref{Eqn:Perna}) to act in all three dimensions and alter the drag timescale (Equation \ref{Eqn:orig_drag}) to account for the changing conductivity. The derivation roughly follows  \citet{Zhu_2005}, although we reframe it in notation consistent with the previous discussion. From Equation \ref{Eqn:BForce}, it is the perpendicular current that is relevant to the magnetic force on the gas.   That current, when moved to the reference frame of the planet, becomes

\begin{equation}
\mathbf{j}_\perp = \sigma_\perp\left(\mathbf{E}_\perp + \frac{\mathbf{u}}{c}\times\mathbf{B} \right) - \sigma_\mathrm{H}\left(\mathbf{E}_\perp + \frac{\mathbf{u}}{c}\times\mathbf{B}\right)\times \mathbf{b}\,\, .
\label{Eqn:jperp}
\end{equation}

\noindent This can then be substituted into Equation \ref{Eqn:BForce}, using $\mathbf{E}_\perp\times\mathbf{B} =\mathbf{E}\times\mathbf{B}$, to arrive at

\begin{align}
 \left(\rho \frac{\partial \mathbf{u}}{\partial t}\right)_\mathrm{magnetic} = & \frac{\sigma_\perp}{c}\left(\mathbf{E}\times \mathbf{B} - \frac{B^2}{c}\mathbf{u}_\mathrm{\perp} \right)\nonumber \\
 & - \frac{\sigma_\mathrm{H}}{c}\left[\left(\mathbf{E}\times \mathbf{B}\right)\times\mathbf{b}- \frac{B^2}{c}\mathbf{u}_\mathrm{\perp}\times\mathbf{b}\right]\,\, .
\end{align}

\noindent The velocity of the field lines in the frame of the computational domain -- the velocity of the frame in which $\mathbf{E}$ vanishes -- is $\mathbf{v}_\mathrm{f}=c\mathbf{E}\times\mathbf{B}/B^2$ which allows the previous equation to be written as
\begin{align}
 \left(\rho \frac{\partial \mathbf{u}}{\partial t}\right)_\mathrm{magnetic} = & -\frac{\sigma_\perp B^2}{c^2}\left(\mathbf{u}_\mathrm{\perp} - \mathbf{v}_\mathrm{f}\right) + \frac{\sigma_\mathrm{H}B^2}{c^2}\left(\mathbf{u}_\mathrm{\perp} - \mathbf{v}_\mathrm{f}\right)\times\mathbf{b}\,\, .
 \label{Eqn:BForce2}
\end{align}

\begin{figure}
    \centering
    \includegraphics{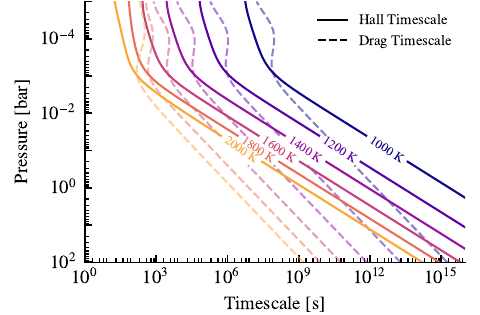}
    \caption{The Hall timescale (solid lines; Equation \ref{Eqn:tauHall}) computed using the Hall conductivity (Equation \ref{Eqn:sigHall}) for a $10\,\mathrm{G}$ magnetic field for a number of different temperatures.  The drag timescales used in this work, also shown in Figure \ref{fig:drag_timescales}, are included as dashed lines for comparison.}
    \label{fig:hall_timescales}
\end{figure}

Taking $v_\mathrm{f}=0$, thus forcing the magnetic field to be stationary in frame of the planet, and introducing the drag timescale $\tau_\mathrm{drag}$ and the Hall timescale $\tau_\mathrm{Hall}$, 

\begin{align}
\tau_\mathrm{drag} & = \frac{c^2\rho}{\sigma_\perp B^2} \,\, ,\label{Eqn:taudrag}\\
\tau_\mathrm{Hall} & = -\frac{c^2\rho}{\sigma_\mathrm{H} B^2} \,\, ,\label{Eqn:tauHall}
\end{align}

\noindent the acceleration on the gas (Equation \ref{Eqn:BForce2}) becomes

\begin{equation}
    \left(\frac{\partial \mathbf{u}}{\partial t}\right)_\mathrm{magnetic} = -\frac{\mathbf{u}_\perp}{\tau_\mathrm{drag}} - \frac{\mathbf{u}_\perp\times\mathbf{b}}{\tau_\mathrm{Hall}}\,\, , 
    \label{eqn:drag3d}
\end{equation}

\noindent which is the governing equation of the extended drag model.   The first term on the right hand side is the generalized drag, which is the focus of this work. The second term, dependent on the Hall timescale $\tau_\mathrm{Hall}$, exerts a force perpendicular to both the bulk velocity and the magnetic field, but only becomes comparable to the drag timescale  when the electrons begin to decouple from the neutral gas and attach to the magnetic field (when $\omega_\mathrm{e}\tau_\mathrm{en}\sim 1$) which, for parameters adopted here, occurs between pressures of 1 mbar and 0.1 mbar. Figure \ref{fig:drag_timescales} shows the drag timescale used in our analysis computed using Equation \ref{Eqn:taudrag} and the timescale computed using the resistivity from \citet{perna_2010a} (Equation \ref{Eqn:etaPerna}).   At high pressures, the two timescales show excellent agreement; however, at $P\sim 10^{-3}\,\mathrm{bar}$, the timescales begin to diverge, with the Perna timescale being an over-estimation of the drag timescale in this low pressure region.  For comparison, the Hall timescale is shown in Figure \ref{fig:hall_timescales}.  Deep in the atmosphere, the Hall timescale exceeds the drag timescale by many orders of magnitude, consistent with being in the Ohmic regime; however, it becomes comparable to the drag timescale at $P\sim 10^{-3}\,\mathrm{bar}$, and ultimately becomes shorter than the drag timescale (compare Figures \ref{fig:drag_timescales} and \ref{fig:hall_timescales}). The Hall timescale becoming shorter than the drag timescale does not, however, indicate that it will play a larger role in the dynamics, as the drag and Hall accelerations are always perpendicular (see Equation \ref{eqn:drag3d}). Within the equatorial jet, for example, the Hall term results in a deflection of the flow with a radius of curvature $\tau_\mathrm{Hall}u_\perp$ that, except in the hottest cases around the substellar point, approaches, if not exceeds, the planetary radius. Consideration of the Hall term would thus be more appropriate for the case of ultra-hot Jupiters, although for these hotter, more ionized planets the drag formalism adopted here may not be suitable and more traditional non-ideal magnetohydrodynamics is presumably a better approach. As a result, in this study, we opt to focus solely on the impact of the generalised drag term.  We note that the impact of the Hall effect in hot Jupiter atmospheres, not in the context of drag models but on a more general, microphysical level, is discussed in \citet{koskinen_2014}.

 We approximate the heating rate due to drag $Q_\mathrm{magnetic} = \mathbf{j}\cdot \mathbf{E}$ using Equation \ref{Eqn:jperp}\footnote{As we take the plasma to be in the frame of the neutrals, this approximation for the heating will become less accurate as the plasma begins to couple to the magnetic field. }, 

\begin{align}
Q_\mathrm{magnetic} & =  \left[\sigma_\perp\left(\mathbf{E}_\perp + \frac{\mathbf{u}}{c}\times\mathbf{B} \right) - \sigma_\mathrm{H}\left(\mathbf{E}_\perp + \frac{\mathbf{u}}{c}\times\mathbf{B}\right)\times \mathbf{b}\right]\nonumber \\
& \,\, \cdot \left(\mathbf{E}_\perp + \frac{\mathbf{u}}{c}\times\mathbf{B}\right) \\
& =  \sigma_\perp \Big|\mathbf{E}_\perp + \frac{\mathbf{u}}{c}\times\mathbf{B}\Big|^2\,\, .
\end{align}

\noindent   Noting that $|\mathbf{A}| = |\mathbf{A}\times \mathbf{b}|$ for a vector $\mathbf{A}$ perpendicular to a unit vector $\mathbf{b}$,  $Q_\mathrm{magnetic}$ can be rewritten as

\begin{align}
Q_\mathrm{magnetic}  & = \sigma_\perp \Big|\mathbf{E}_\perp \times \mathbf{b} - \frac{B}{c}\mathbf{u}_\perp \Big|^2 = \frac{\sigma_\perp B^2}{c^2}\big|\mathbf{u}_\perp - \mathbf{v}_\mathrm{f}\big|^2 \\
 & = \frac{\rho(|u_\perp|^2+|v_\perp|^2+|w_\perp|^2)}{\tau_\mathrm{drag}} \,\, ,\label{Eqn:heating}
\end{align}

\noindent where again $\mathbf{v}_\mathrm{f} = 0$ has been assumed.   Equation \ref{Eqn:heating} is the same drag heating rate used in \citet{rauscher_2013} and \citet{beltz_2022}, generalized to include vertical and meridional drag.  The Hall conductivity terms, and thus all Hall timescale terms, vanish, and as a result all magnetic heating is due to drag.  

To account for the magnetic field geometry, we first assume that the planet's magnetic field is a dipole aligned with the rotational axis of the planet, which also corresponds to the coordinate axis:  

\begin{align}
        B_{\lambda} & = 0 \nonumber\,\, ,\\
        B_{\theta} & = \frac{B_\mathrm{ref}}{2}\left(\frac{r_\mathrm{ref}}{r}\right)^3\cos\theta \,\, ,\nonumber\\
        B_{r} & = B_\mathrm{ref}\left(\frac{r_\mathrm{ref}}{r}\right)^3\sin\theta\,\, ,
        \label{Eqn:B}
\end{align}

\noindent where $B_\mathrm{ref}$ is the magnetic field strength at the poles at a reference radius of $r=r_\mathrm{ref}$.  For simplicity, we have opted to use a radius corresponding to the midpoint of the computational domain in our fiducial case ($r_\mathrm{ref}=9.5\times 10^7\,\mathrm{m}$).   With the magnetic field specified,  the component of the velocity perpendicular to the local magnetic field, $\mathbf{u_\perp} = (u_{\perp},v_{\perp},w_{\perp})$, is given by $\mathbf{u}_\perp = -\left(\mathbf{u}\times\mathbf{b}\right)\times \mathbf{b}$, or, in component form,

\begin{align}
    u_{\perp} & = u \label{Eqn:perp_u}\,\, ,\\
    v_{\perp} & = \frac{4v\sin^2\theta + 2w\sin\theta\cos\theta}{1+3\sin^2\theta} \label{Eqn:perp_v}\,\, ,\\
    w_{\perp} & = \frac{w\cos^2\theta+2v\sin\theta\cos\theta}{1+ 3\sin^2\theta } \label{Eqn:perp_w}\,\, .
\end{align}

We first note that when considering only drag in the zonal direction (i.e., $v_{\perp}$ and $w_{\perp}$ are taken to be zero), this reduces to the model of \citet{perna_2010a} (see also Equation \ref{Eqn:Perna}), modulo differences in how the drag timescale is computed.  Second, in a regime where $|w| \ll |v|$, the wind experiences a meridional drag as it moves poleward due to the magnetic dipole geometry (see Equation \ref{Eqn:perp_v}), and the vertical velocity experiences an acceleration largely dependent on the meridional velocity (see Equation \ref{Eqn:perp_w}).  

\subsection{Model Limitations}
\label{Sec:Limitations}

While the derivation in the previous section generalizes previous drag models and addresses some of the underlying assumptions, it does not address all the assumptions as it remains a simple drag model attempting to account for complex magnetic effects within the atmosphere. The fundamental, unaddressed assumption is that the field largely remains static and the assumed field configuration is as least approximately representative of the configuration one would find were the equations of magnetohydrodynamics to be solved directly. This is essential to the reduction of all magnetic effects to a magnetic drag term.  

For the sake of simplicity, the derivation presented assumes a dipolar magnetic field, although it is not assured that an initially dipolar field remains so.  The ability of atmospheric winds to deform the magnetic field is often expressed in terms of the magnetic Reynolds number, $R_\mathrm{m}$, which attempts to estimate the relative importance of the advective and diffusive terms in the magnetic induction equation.  While generally $R_\mathrm{m} \propto LV/\eta$, for some characteristic length scale $L$, velocity $V$, and resistivity $\eta$, we specifically adopt the pressure scale height $H$, horizontal wind velocity $u_\mathrm{horiz}$, and perpendicular resistivity $\eta_\perp$ yielding

\begin{equation}
    R_\mathrm{m} = \frac{4\pi H u_\mathrm{horiz}}{c^2\eta_\perp}\,\, ,
    \label{Eqn:Rm}
\end{equation}

\noindent where the factor of $4\pi c^{-2}$ accounts for our choice of definition of the resistivity.   Regions with $R_\mathrm{m} \gg 1$ are assumed to be better coupled with the wind and able to advect the field while in regions with $R_\mathrm{m} \ll 1$ the gas is capable of moving through the field while not necessary inducing large deformation of the field lines, the latter being the region where the Lorentz forces on the gas can best be approximated as a drag term.\footnote{ Caution should be exercised in over-interpreting $R_\mathrm{m}$ in favour of drag models, however, as choices in characteristic velocity $U$ can yield undue confidence in the strict validity of the drag approximation.  For example, we assume $U=u_\mathrm{horiz}$ here.  If the drag is sufficiently large due to small resistivities, horizontal motions may be suppressed resulting in a small Reynolds numbers.   The suppression of the velocities is due to the underlying assumption of a static field, and were the evolution of the field to be followed, larger velocities may result as the magnetic field can now be advected.   Similarly, in a zonal drag model, assuming the characteristic velocity is the zonal velocity (i.e., $U=u$ in the nomenclature here) might result in a smaller Reynolds number as strong zonal drag results in meridional velocities which aren't considered in the adopted $U$, despite the model reaching the point where the advection of the field needs to be considered. }  As the drag model is applied to increasingly irradiated planets, the strict validity of the drag model will break down, first on the dayside as the temperatures increase and then eventually on the nightside, depending on the efficiency of day-night heat transport.     

The accuracy of the magnetic drag approximation has been investigated by \citet{rogers_2014} and \citet{rogers_2014b}. In their simulations, the equations of non-ideal magnetohydrodynamics are solved directly, assuming a horizontally uniform resistivity based on a reference temperature profile.    In comparing the Lorentz forces in their ``hot'' simulation of HD~209458~b to prescribed drag force from \citet{perna_2010a} with a fixed magnetic field strength, \citet{rogers_2014} find an order-of-magnitude over-estimation in the force on the atmosphere from the drag scheme.   This is likely due to the induced toroidal field within the magnetohydrodynamic model which winds the magnetic field around the planet.  The field is thus increasingly horizontal, especially at the equator, and the motions of the jet are able to flow along the field lines instead of across them as is assumed by the static dipolar geometry of the magnetic drag model.  The induced toroidal field in the model of \citet{rogers_2014} is shown in their Figure 3, although Figure 3 of \citet{rogers_2014b} provides more illustrative examples of the possible field deformations.

This is especially relevant for our discussion of magnetic drag models as the assumption of solely meridional currents causes the Lorentz forces to vanish at the equator.  In the generalized drag model presented here, that assumption is no longer made, allowing for the possibility of equatorial drag, and as the assumed field geometry is that of an aligned dipole, increased equatorial drag will, in fact, occur.  If instead the magnetic field is primarily toroidal near the equator, as is the case in many of the simulations of \citet{rogers_2014} and \citet{rogers_2014b}, the Lorentz forces will be negligible resulting in the \citet{perna_2010a} drag prescription providing better agreement with the magnetohydrodynamic simulations, although the caveat made in \citet{rogers_2014} that the drag models over-estimate the Lorentz forces remains.  

While the magnetohydrodynamic simulations of \citet{rogers_2014} and \citet{rogers_2014b} provide many insights about the limitations of magnetic drag models, assumptions made within them warrant their own discussion.   To make the solution of the induction equation tractable, the resistivity is assumed to be horizontally uniform.  This avoids extremely large resistivities on the nightside where the ionization fractions are extremely small but, as a consequence, artificially inflates the coupling of the magnetic field to the atmosphere on the nightside\footnote{The use of a temperature from a background or average reference state similarly results in an overestimation of the resistivity on the dayside around the hot spot and thus underestimates the coupling there, although this does not introduce numerical challenges.}.   This artificial nightside coupling may improve the ability of the wind to shape the field around the planet (i.e., induce a toroidal component to the magnetic field), thus over-estimating the impact of the toroidal field.  Conversely, the hydrodynamic models of \citet{rogers_2014} and \citet{rogers_2014b} consistently show slower mean zonal jet speeds compared to GCM models.  If the wind speeds are being underestimated, a magnetohydrodynamic model able to reproduce the hydrodynamic limit seen in GCMs might have larger toroidal fields if the changes in the hydrodynamic behaviour translate to changes in the magnetohydrodynamic behaviour. It should also be noted that the models of \citet{rogers_2014} find the largest Lorentz forces to be on the nightside of the planet where the resistivity is over-estimated, thus potentially over-estimating the impact of magnetic effects.  Future modelling that properly models both the resistivity and the large temperature differences between the day and nightsides of hot Jupiters is required before these questions will be resolved.

\section{Simulations}
\label{Sec:Simulations}
We have generalized the magnetic drag models to investigate how relaxing the assumption of solely zonal drag {\bf and meridional currents} and considering the full geometry of the magnetic field will impact the atmospheric dynamics compared to both non-magnetic models and models that use only zonal drag.   This was accomplished through three-dimensional simulations using the Met Office's {\sc Unified Model} GCM modified to include magnetic drag. The details of these model and the parameter study are outlined below.


\subsection{The Unified Model}
\label{SubSec:UM}

The {\sc UM} solves the fully-compressible, deep-atmosphere, non-hydrostatic Euler equations \citep[][]{mayne_2014a,wood_2014}. While it has been used to model hot Jupiters \citep{mayne_2014a,mayne_2017} and mini-Neptunes \citep{drummond_2018,mayne_2019}, this work constitutes the first attempt to include magnetic drag.  Radiative transfer is calculated using the {\sc Socrates} radiative transfer code based on \citet{edwards_1996} using pseudo-spherical approximation outlined in \citet{jackson_2020}. Gas-phase opacity sources are \ce{H2O}, CO, \ce{CH4}, \ce{NH3}, Li, Na, K\footnote{We do not consider the ionization of Na and K in the opacity model as it is computed separately in the resistivity calculation.  This will result in an overestimation of the total opacity due to these species.}, Rb, Cs, and \ce{H2}-\ce{H2} and \ce{H2}-He collision induced absorption (CIA) with the opacities computed using the correlated-k method. We use the analytic chemistry of \citet{burrows_1999}, extended to include alkali metals, assuming solar elemental abundances taken from \citet{lodders_2009}.

\begin{table}
\caption{Common Simulation Parameters}
\label{Tbl:Common}
\begin{tabular}{lcc}
\hline
  & Value & Units\\
\hline
{\em Grid and Timestepping} \\
Longitude Cells & 144 &\\
Latitude Cells & 90 &\\
Vertical Layers & 80 & \\
Hydrodynamic Timestep & 30 & s \\
Simulation Length & 1000 & Earth days \\
\\
{\em Radiative Transfer} \\
$F_{\star,\mathrm{HD209458b}}$ (at 1 AU) & 2054.73 & $\mathrm{W\, m^{-2}}$ \\
Radiation Timestep & 150 & s \\
\\
{\em Damping and Diffusion} \\
Damping Profile & Horizontal &\\
Damping Coefficient & 0.15 &\\
Damping Depth ($\eta_s$) & 0.8 & \\
\\
{\em Planet}\\
Magnetic Field ($B_\mathrm{ref}$) & 10 & G \\
Intrinsic Temperature ($T_\mathrm{int}$) & 100 & K \\
Initial Inner Boundary Pressure & 200 & bar \\
Inner Boundary Radius ($R_\mathrm{inner}$) & $9.0\times 10^7$ & m \\
Gravitational acceleration at $R_\mathrm{inner}$ ($g$)  & 10.79 & $\mathrm{m\,s^{-2}}$\\
Semi-major axis ($a$) & $4.747\times 10^{-2}$ & AU \\
Ideal Gas Constant ($R$) & 3556.8 & $\mathrm{J\,kg^{-1}\,K^{-1}}$ \\
Specific Heat Capacity ($c_\mathrm{P}$) & $1.3\times 10^4$ &  $\mathrm{J\,kg^{-1}\,K^{-1}}$ \\
Angular velocity ($\Omega$) & $2.06\times 10^{-5}$ & $\mathrm{s^{-1}}$ \\
\hline
\end{tabular}
\end{table}

\begin{table}
\caption{Computational Domain Heights}
\label{Tbl:Domain}
\begin{tabular}{lc}
\hline
 Stellar Flux & Domain Height (m) \\
\hline
$F_{\star,\mathrm{HD209458b}}$ & $1.0\times 10^7$ \\
1.5$F_{\star,\mathrm{HD209458b}}$ & $1.1\times 10^7$ \\
2$F_{\star,\mathrm{HD209458b}}$ & $1.1\times 10^7$ \\
3$F_{\star,\mathrm{HD209458b}}$ & $1.2\times 10^7$  \\
4$F_{\star,\mathrm{HD209458b}}$ & $1.3\times 10^7$  \\
\hline
\end{tabular}
\end{table}
\subsection{The Parameter Study}
\label{SubSec:ps}
To understand to what extent the expanded drag model is relevant, we opt to model the hot Jupiter HD\,209458\,b as it represents a relatively cool target ($T_\mathrm{eq} \sim 1290$ K; \citealt{xue_2024}) where magnetic drag is expected to act as a perturbation to the flow \citep{rauscher_2013} as opposed to dramatically altering the flow \citep[e.g.,][]{beltz_2022}. We then, holding all other planetary parameters fixed, perform simulations with increased levels of instellation to explore how the flow changes as the atmosphere becomes hotter and the dayside begins to experience increased magnetic drag.   This also corresponds to simulations with the dayside satisfying $R_\mathrm{m}\ll 1$ in the fiducial case with the the strict applicability of the drag model breaking down  ($R_\mathrm{m} > 1$) on the dayside of the most irradiated cases.  

The common simulation parameters adopted for all simulations are shown in Table \ref{Tbl:Common}, with instellations of $1\times$, $1.5\times$, $2\times$, $3\times$, and $4\times$ the instellation of HD\,209458\,b ($F_{\star,\mathrm{HD209458b}}$).  As increased instellation results in a more extended dayside and as the UM uses a height-based vertical grid (instead of a pressure-based grid), we vary the domain height with the instellation (see Table \ref{Tbl:Domain}) so that the maximum pressure on the outer boundary never exceeds a few $\times 10^{-5}$ bar, and remains approximately comparable between simulations.

We performed our simulations with a reference magnetic field strength of $B_\mathrm{ref}=10\,\mathrm{G}$.  Due to the model considering the magnetic geometry, the magnetic field strength varies with position across the computational domain (see Equation \ref{Eqn:B}), reaching a maximum of $11.76\,\mathrm{G}$ at the inner boundary at the poles and a minimum of $4.286$ to $3.923\,\mathrm{G}$ at the outer boundary at the equator, depending on the radial extent of the computational domain.

\section{Results}
\label{Sec:Results}

In this section we present the results of the suite of simulations.   We begin with the simulations of the fiducial HD\,209458\,b, and move on to cases with increased temperatures and magnetic drag. 
\subsection{The Fiducial HD\,209458\, b}

\begin{figure*}
    \centering
    \includegraphics{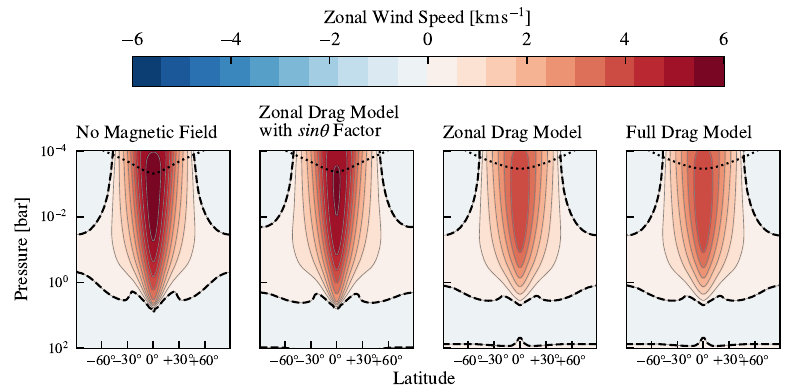}
   \caption{The mean zonal wind for the fiducial ($F_\star=F_{\star,\mathrm{HD209458b}}$) case with no magnetic field (leftmost panel), zonal drag with the inclusion of the $\sin\theta$ factor (second panel from the left), zonal drag (third panel from the left), and the full drag model (rightmost panel). The thick black dashed line denotes the boundary between super-rotation and counter-rotation.  The thick dotted line indicates where the gas begins to intersect the vertical sponge.}
    \label{fig:sc1zonalwind}
\end{figure*}

\begin{figure*}
    \centering
    \includegraphics{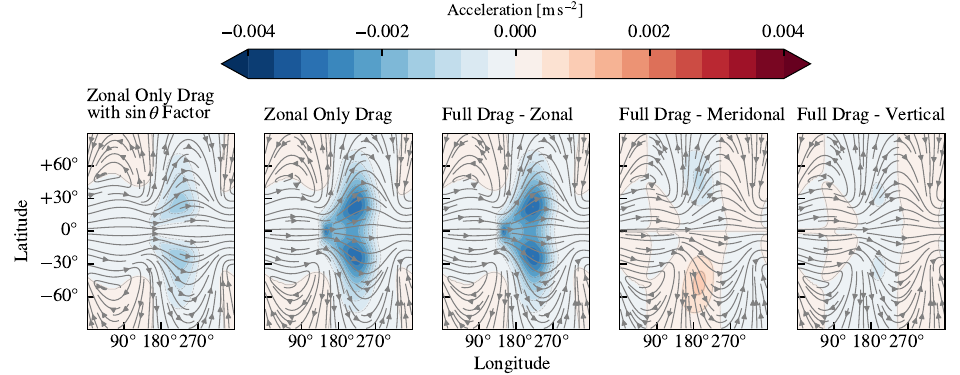}
   \caption{The accelerations due to magnetic drag for three $F_\star=F_{\star,\mathrm{HD209458b}}$ simulations at a pressure of 1 mbar.  The leftmost two panels show the acceleration for the zonal drag models with and without the $\sin\theta$ factor while the remaining three show the individual components of the acceleration for the full drag model.}
    \label{fig:sc1drag}
\end{figure*}

Before examining how the magnetic drag impacts the dynamics of increasingly irradiated atmospheres, we first detail the fiducial case of HD\,209458\, b.   We simulate four different cases each with an instellation of $F_\star = F_{\star,\mathrm{HD209458b}}$ --  a hydrodynamic case without drag, a zonal drag case which includes a factor of $\sin\theta$ in the drag timescale which mimics the assumption of solely meridional currents, a zonal drag case without the factor of $\sin\theta$ which we will just refer to as the zonal drag case, and a full drag case -- to examine how the zonal and full drag models compare in a part of parameter space where they are expected to be only a perturbation to the flow.    For the zonal drag model which includes the $\sin\theta$ factor, we take the drag timescale used in the other models, Equation \ref{Eqn:taudrag}, and divide it by the aforementioned factor, 

\begin{equation}
\tau_\mathrm{drag,with\,sin\theta} = \frac{c^2\rho}{\sigma_\perp B^2\sin\theta}\,\, .
\label{Eqn:dragwithsin}
\end{equation}

\noindent While this timescale is closer in form to the drag timescale of \citet{perna_2010a}, it differs from the Perna drag timescale (Equation \ref{Eqn:orig_drag}) used in other works in that it depends on the perpendicular conductivity instead of the parallel conductivity (or, equivalently, the scalar resistivity).  We make this choice to isolate the effect of the assumption of meridional currents through the inclusion of the $\sin\theta$ factor and to better compare its impact to the other drag models.

The mean zonal windspeeds for each case are shown in Figure \ref{fig:sc1zonalwind}.  The non-magnetic case has a peak mean zonal windspeed of 5.783\,$\mathrm{km\,s^{-1}}$ while the zonal drag model that includes the factor of $\sin\theta$ has a slightly slower peak mean windspeed of 5.499\,$\mathrm{km\,s^{-1}}$.  The zonal drag model without the factor of $\sin\theta$ and full drag model have peak mean zonal windspeeds of 4.021 and 4.026 $\mathrm{km\,s^{-1}}$, respectively.  The full drag model thus results in only a minor change in windspeeds relative to  our fiducial zonal drag model that does not include the $\sin\theta$ factor.  The largest difference is seen between the two zonal drag models where the removal of the $\sin\theta$ factor decreases the peak zonal windspeed from $5.499$ to $4.021$ $\mathrm{km\,s^{-1}}$.  It is thus the breakdown of the assumption of predominantly meridional currents, which is the source of the $\sin\theta$ factor, that has the largest impact in cooler, less irradiated atmospheres.  This shows that zonal drag models may still be appropriate for the cooler hot Jupiters if the factor of $\sin\theta$ is omitted\footnote{ We do not make any specific claims regarding the degree to which other works may be impacted by this, as their simulation setups can be significantly different from what we have used here.  For example, the model of HD\,209458\, b in \citet{rauscher_2013} uses dual grey radiative transfer and parameters that result in a hotter atmosphere than those we find for our fiducial HD\,209458\,b case.}.  The origins of these differences can  be appreciated by comparing the accelerations between the various models (see Figure \ref{fig:sc1drag}). At, for example, 1 mbar, the peak zonal drag in the model with the $\sin\theta$ factor is only 12\% of the peak zonal drag in the model without the $\sin\theta$ factor, illustrating the impact of the change in drag timescales, while  the zonal and full drag models differ by only 0.62\%. The peak meridional drag in the full drag model at 1 mbar (see Figure \ref{fig:sc1drag}, 3rd panel), is 3.5 times smaller and relatively localized on the dayside at mid-latitudes, due to the cooler polar temperatures and largely zonal winds, limiting the impact of its inclusion in the model.   The vertical accelerations (Figure \ref{fig:sc1drag}, rightmost panel) are even smaller by comparison.  This increased suppression of the equatorial jet due to the removal of the assumption of purely meridional currents remains conditional on the assumption that the field remains primarily poloidal.  If the equatorial jet is capable of winding the field, creating a toroidal component, the equatorial suppression of the jet will be reduced as flow of the jet will be along, not across, the now toroidal field.  This can be seen in the {\em hot}  model of HD~209458~b in \citet{rogers_2014} where the Lorentz forces in the magnetohydrodynamic model vanish at the equator (see their Figure 2). As discussed in Section \ref{Sec:Limitations}, these models do assume a horizontally uniform magnetic resistivity which increases the nightside coupling \citep[as do ][]{rogers_2014b,batygin_2013} and thus facilitates the winding and generation of a toroidal field.  They do, nonetheless, provide insight on how the model may break down. Insofar as the goal here is to compare the assumptions of specifically magnetic drag models, and given the degree of agreement between the zonal and full drag models, except for differences induced by the inclusion of the $\sin\theta$ factor, we do not compare these less irradiated models any further.

\subsection{The Dynamics of Hotter Atmospheres}

\begin{figure*}
    \centering
    \includegraphics{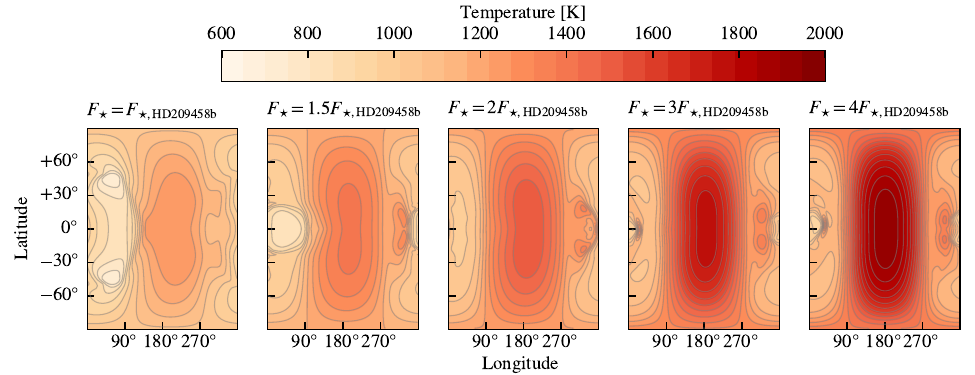}
    \caption{Temperature at a pressure level of 1 mbar for each of the full drag models.}
    \label{fig:horiztemp}
\end{figure*}

\begin{figure*}
    \centering
    \includegraphics{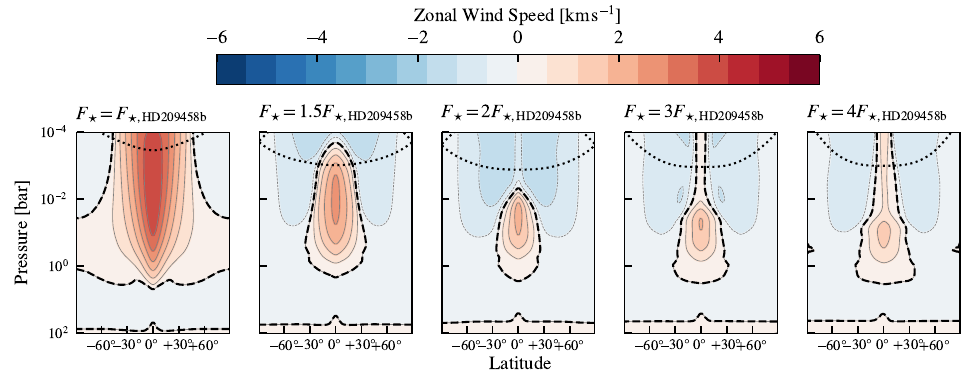}
    \caption{The zonal-mean wind speed for a $B_\mathrm{ref}=10\, \mathrm{G}$ magnetic field for increasing levels of stellar flux. Positive values indicate super-rotation. The thick black dashed line denotes the boundary between super-rotation and counter-rotation.  The thick dotted line indicates where the gas begins to intersect the vertical sponge. The narrow equatorial super-rotating feature seen at $P< 10^{-2}\,\mathrm{bar}$ in the leftmost two panels is not a narrow super-rotating jet but instead is due to the localized velocities at the anti-stellar point where the flows converge (see Figure \ref{fig:horizwind}).}
    \label{fig:zonalwind}
\end{figure*}

\begin{figure*}
    \centering
    \includegraphics{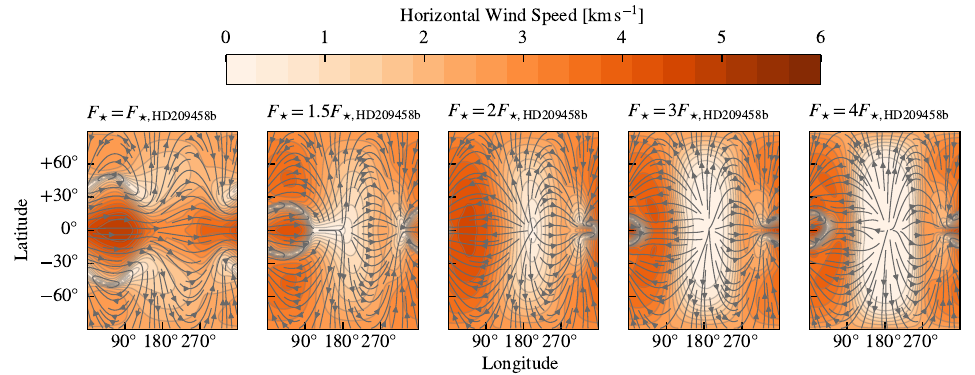}
    \caption{The horizontal wind speed at a pressure level of 1 mbar. The overplotted streamlines indicate the direction of the direction of the horizontal velocity. }
    \label{fig:horizwind}
\end{figure*}

\begin{figure*}
    \centering
    \includegraphics{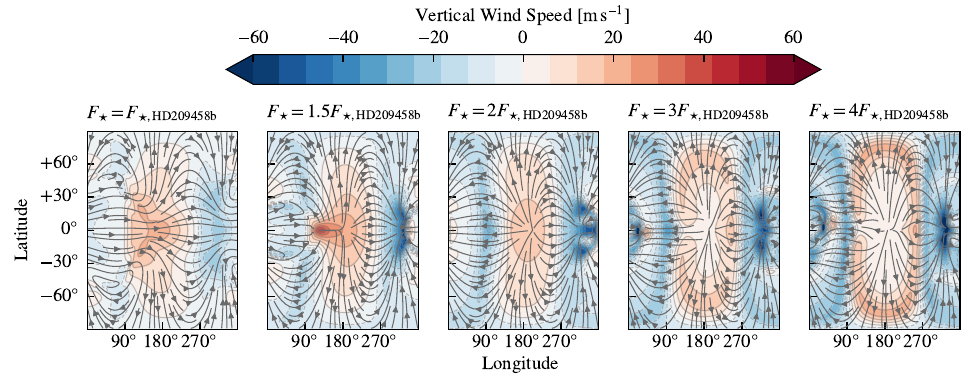}
    \caption{The vertical velocity at a pressure level of 1 mbar.  Positive (red) colours indicate upward motions while negative (blue) colours indicate downward motions. The overplotted streamlines indicate the direction of the horizontal velocity.  }
    \label{fig:vertwind}
\end{figure*}
To understand how the dynamics will change as the atmospheric temperatures get hotter, we increase the instellation, as discussed in Section \ref{SubSec:ps}.  The resulting temperatures for the full drag model are shown in Figure \ref{fig:horiztemp}, and the mean zonal winds are shown in Figure \ref{fig:zonalwind}.  For the fiducial HD\,209458\,b case ($F_\star=F_{\star,\mathrm{HD209458b}}$), a zonal jet forms, peaking at $\sim 4\,\mathrm{km\,s^{-1}}$, slower than a model without magnetic drag but qualitatively similar in the flow geometry as was seen in the previous section and similar to previous simulations of HD\,209458\,b using the UM \citep[e.g.,][]{zamyatina_2023}.  In this least irradiated case, the super-rotating jet extends around the entire equatorial region, as can be observed in the leftmost panel of  Figure \ref{fig:horizwind}.   As the atmosphere gets hotter with increased instellation, the super-rotating jet in the upper atmosphere begins to be suppressed, transitioning instead to a slow diverging flow from the substellar point (see the horizontal velocities at 1 mbar in Figure  \ref{fig:horizwind}).  The super-rotating jet is not suppressed uniformly with altitude, and it does persist deeper in the atmosphere although the dayside velocities are dramatically reduced, with a peak mean zonal velocity of $\sim 1\,\mathrm{km\,s^{-1}}$ in the most irradiated case. and the deep jet may similarly be suppressed as temperatures increase further.    The suppression of the jet, seen both in previous magnetic drag models \citep[e.g.,][]{beltz_2022} and in the models presented here, may alter the deep atmosphere adiabat as the jet is responsible for the advection of potential temperature into the deep atmosphere, as discussed in \citet{tremblin_2017} and \citet{sainsbury-martinez_2019}.  

\begin{figure*}
    \centering
    \includegraphics{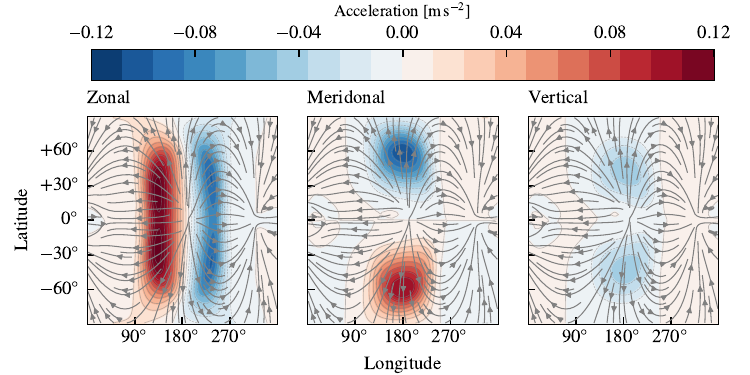}
    \caption{The zonal (left panel), meridional (center panel), and vertical (right panel) accelerations due to magnetic drag for the $F_\star=3F_{\star,\mathrm{HD209458b}}$ case at a pressure level of 1 mbar. The overplotted streamlines indicate the horizontal velocity.}
    \label{fig:dragF3}
\end{figure*}

The dayside diverging flow seen at 1 mbar in Figure \ref{fig:horizwind} is extremely slow moving, unlike in the zonal drag models of \citet{beltz_2022} where the jet transitions to being meridional.   This is due the inclusion of meridional and vertical drag creating a quasi-static substellar region, and while the horizontal velocities increase near the terminators and across the poles, the upwelling of gas from deeper in the atmosphere is not at the equator but in an annulus around this  quasi-static region (see Figure \ref{fig:vertwind}). This can be seen in the accelerations on the gas due to magnetic drag, shown for the $F_\star = F_{\star,\mathrm{HD209458b}}$ and $F_\star = 3F_{\star,\mathrm{HD209458b}}$ cases in Figures \ref{fig:sc1drag} (rightmost 3 panels) and \ref{fig:dragF3}, respectively.   For the less irradiated case, the drag is $\sim 30$ times less in magnitude compared to the more irradiated case, and unable to suppress the formation of the equatorial jet, and the zonal component of the drag dominates due to the motions remaining primarily zonal.   In the more irradiated case (Figure \ref{fig:dragF3}), the zonal drag increases due to the hotter temperatures and thus shorter drag timescales, suppressing the formation of the equatorial jet.  In a purely zonal drag model, this would result in meridional flow over the pole; however, motions towards the poles, especially at the mid-latitudes ($\sim \pm 60^\circ$), experience drag (Figure \ref{fig:dragF3}, middle panel) and an acceleration downward (Figure \ref{fig:dragF3}, right panel), as the field lines of the magnetic dipole curve inward.   This behaviour is qualitatively similar to what is seen in some magnetohydrodynamic models of atmospheric escape , such as \citet{trammell_2014}, \citet{khodachenko_2015}, or \citet{presa_2024}, where the equatorial gas becomes trapped by closed field lines, preventing the launching of an escaping wind, and resulting in an equatorial {\em dead zone}.\footnote{This terminology may lead to some confusion.  Within models of atmospheric escape, the dead zone is due to well-coupled gas trapped by the planetary magnetic field, unable to escape.  This is contrasted with the dead zone within the protoplanetary disks which are not turbulent near the midplane due to a lack of coupling to the magnetic field.}  It is not, on the other hand, seen in the magnetohydrodynamic simulations of \citet{rogers_2014} and \citet{rogers_2014b}.  

The hot quasi-static region near the substellar point may, however, simply be a consequence of a breakdown of the model and the underlying assumption of a static field.   The substellar atmosphere is sufficiently ionised that it is coupled to the magnetic field; however, it is, by assumption, unable to advect the field along with the flow.  While a well-coupled yet static dayside could still form if the magnetic pressure was greater than the gas pressure (i.e., the plasma $\beta$ $=P/(B^2/8\pi)$, the ratio of gas and magnetic pressures, being greater than unity), on the dayside, this only occurs at the upper boundary of the simulation and is significantly less than unity in the rest of the dayside computational domain.  As such, one would expect in these regions where the gas is well coupled to the magnetic field that the gas should be able to move without significant opposition due to magnetic pressure, making it unclear as to whether the magnetic field in these regions can suppress the flow in the manner shown in the simulations, although the degree to which it departs from this behaviour is unclear. 

\begin{figure*}
    \centering
    \includegraphics{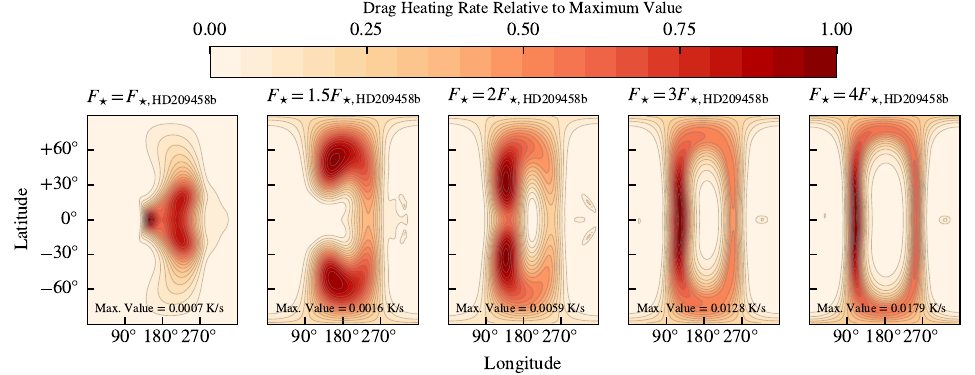}
    \caption{The heating due to magnetic drag at 1 mbar.  Due to the differing peak heating rates between simulations, the colour scales have been normalized to the maximum value in the plot with the peak value indicated at the bottom of the panel.  }
    \label{fig:dragheat}
\end{figure*}

\begin{figure*}
    \centering
    \includegraphics{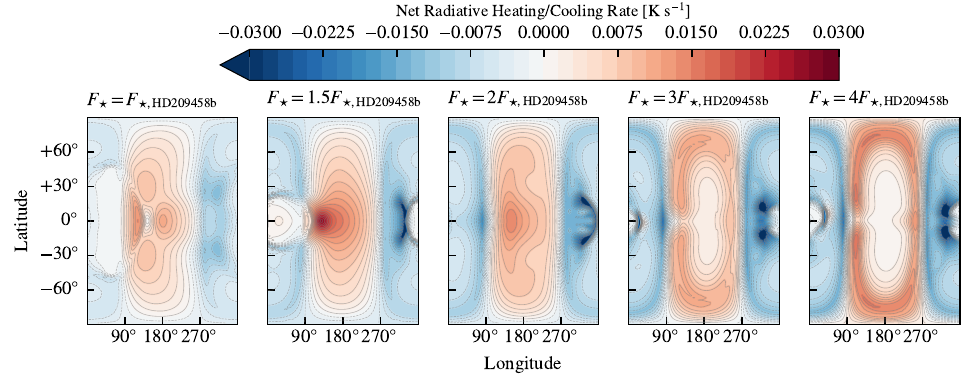}
    \caption{The net radiative heating and cooling in the atmosphere at 1 mbar. }
    \label{fig:radheat}
\end{figure*}

The magnetic heating within the atmosphere, shown in Figure \ref{fig:dragheat} transitions from primarily contributing at the equator in the least irradiated case to forming an annulus, following the upwelling and subsequent flow across the terminator, in the most irradiated cases.  While the heating is localized, especially for the most irradiated case, in the upper atmosphere ($P\sim 10^{-3}\,\mathrm{bar}$), the average heating becomes comparable to the radiative cooling, and is likely to impact the thermal balance in those parts of the atmosphere. The impact of the drag can also be seen in the net radiative heating rates (Figure \ref{fig:radheat}) where the substellar point, being nearly static, is in radiative equilibrium while the annulus surrounding it is out of radiative equilibrium due to a combination of magnetic drag and transport of cool gas from deeper in the atmosphere.

\begin{figure*}
    \centering
    \includegraphics{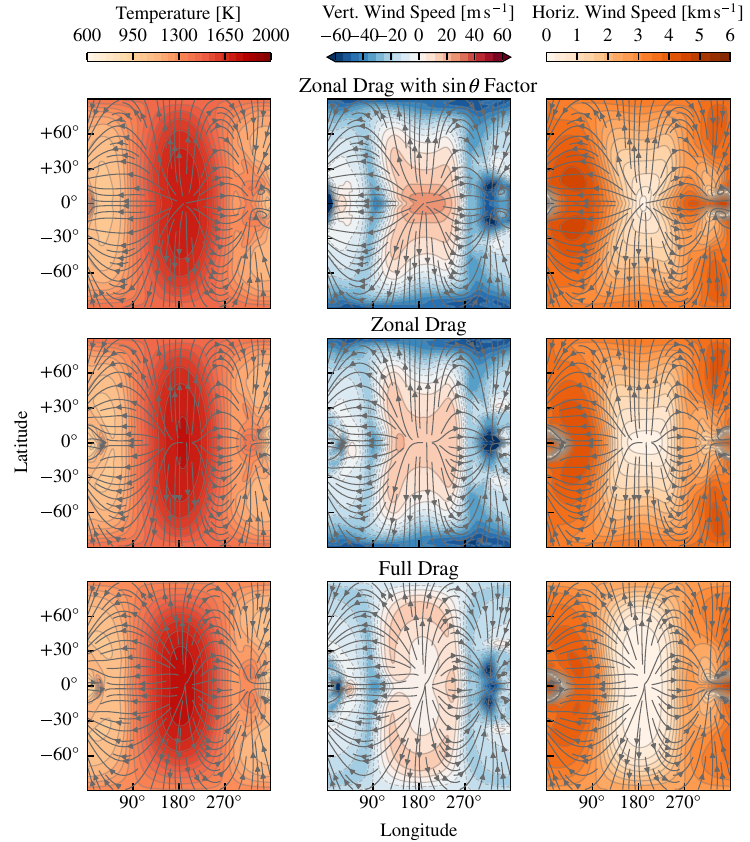}
    \caption{The temperature (left column), vertical wind speed (centre column), and horizontal wind speed (right column) at 1 mbar for simulations with $F_\star=3F_{\star,\mathrm{HD209458b}}$. The top row shows a simulation with
    zonal drag including a $\sin\theta$ factor in the drag timescale, the middle row shows a simulation without the $\sin\theta$ factor in the drag timescale, and the bottom row shows the full drag model.   The bottom row duplicates panels from previous plots, but are included here for ease of comparison.}
    \label{fig:zonaldragcomp}
\end{figure*}

\begin{figure*}
    \centering
    \includegraphics{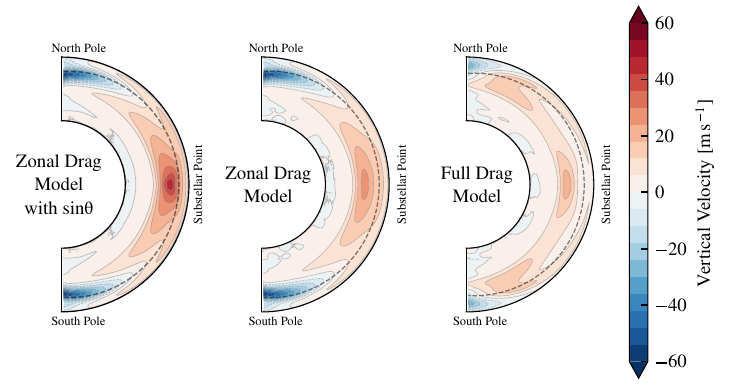}
    \caption{A slice from the rotational axis through the substellar point for the zonal drag models with the $\sin\theta$ factor (left panel) and without (centre panel), and full drag model (right panel) showing the vertical velocity $w$.  The dashed line indicates a pressure of 1 mbar, the pressure level shown in previous figures. The radial scale is exaggerated for clarity.}
    \label{fig:ss_w}
\end{figure*}

\subsection{Comparison with the Zonal Drag Model}

To illustrate better how the zonal drag model differs from the full drag model in the more irradiated, hotter cases, we run  two additional simulations with an instellation of $F_\star=3F_{\star,\mathrm{HD209458b}}$ applying drag only in the zonal direction, one simulation with the inclusion of a $\sin\theta$ factor in the drag timescale (Equation \ref{Eqn:dragwithsin}) and one without. These will still differ from other zonal drag models \citep[e.g.,][]{rauscher_2013} in that we compute the drag timescale using Equations \ref{Eqn:dragwithsin} and  \ref{Eqn:taudrag}, respectively, and we also compute the magnetic field strength using Equation \ref{Eqn:B} instead of using a constant magnetic field strength.

\begin{figure*}
    \centering
    \includegraphics{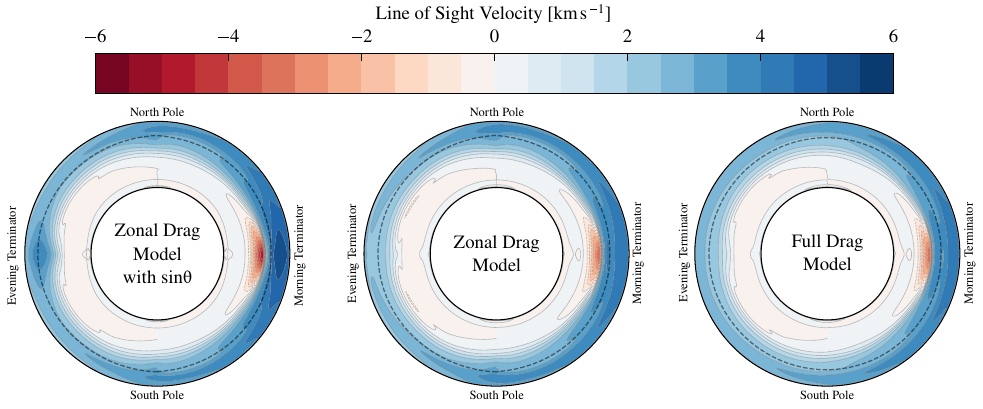}
    \caption{The line-of-sight velocity along a slice through the terminator for the $F_\star=3F_{\star,\mathrm{HD209458b}}$ for both the zonal drag model (left) and the full drag model (right). Negative (red) colours represent motions towards the star while positive (blue) colours represent motions away from the star. The dashed line indicates a pressure of 1 mbar, the pressure level shown in previous figures.  Note that the radial extent of the computational domain has been exaggerated relative to the size of the interior for clarity.  The velocities have not been corrected to include the solid-body planetary rotation.}
    \label{fig:termslice}
\end{figure*}

The temperature, vertical and horizontal velocities for each of the simulations are shown in Figure \ref{fig:zonaldragcomp}.   Unlike in the full drag cases where there exists a quasistatic region around the substellar point with the dayside updraft surrounding in it in an annulus, the updraft in the zonal drag simulations originates at the substellar point with the gas accelerating towards the poles, as has been observed in the simulations of \citet{beltz_2022}.  This is especially prominent in the simulation which includes the $\sin\theta$ factor as, in addition to the meridional flow that forms, there is less zonal drag at the equator, allowing for increased zonal flow. This difference also results in changes near the poles where in the zonal models there exists a fast ($\sim 60\,\mathrm{m\,s^{-1}}$) downdraft but as the full drag models result in the updraft being moved poleward, this downdraft is suppressed.  The differences in the vertical flow can be seen in Figure \ref{fig:ss_w} which shows the vertical velocities through the substellar point.   The equatorial updraft persists deeper in the atmosphere for the full drag model, consistent with the deeper equatorial jet; however, separate updrafts near the poles inhibit the downdrafts seen in the zonal drag models. With the dynamics on the dayside around the substellar point altered, the flow across the terminator, seen in Figure \ref{fig:termslice}, is changed in two ways. First, the velocities at the poles slow due to the inclusion of meridional drag. Second, the equatorial velocities are reduced by the increased equatorial drag due to the removal of the $\sin\theta$ factor in the drag timescale. In the zonal model with the $\sin\theta$ factor included, the line-of-sight velocities on the terminator range from 5.48 $\mathrm{km\,s^{-1}}$ towards the star to $4.79\,\mathrm{km\,s^{-1}}$ away from the star. In the full drag model, these are reduced to $4.25$ and $3.34\,\mathrm{km\,s^{-1}}$, respectively.

\begin{figure}
    \centering
    \includegraphics{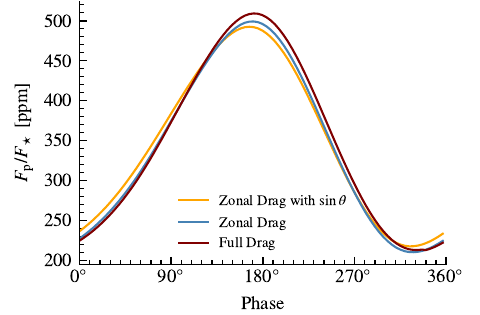}
    \caption{The emission between 2 and 5 microns for zonal drag models with the $\sin\theta$ facor (orange) and without it (blue), as well as full drag model (red), each with $F_\star = 3F_\mathrm{\star,HD209458b}$. Due to the slightly hotter atmosphere near the substellar point, the phase curve peaks $\sim 17$ and $\sim 10\,\mathrm{ppm}$ higher in the full drag model relative to the zonal drag models with and without the $\sin\theta$ factor, respectively.} 
    \label{fig:phasecurve}
\end{figure}

The suppression of dayside meridional motions also results in a slightly warmer ($\sim 50\,\mathrm{K}$) hot spot in the full drag model.   This translates to a modest increase in dayside flux seen in the phase curve ($\sim 17$ and $\sim 10\,\mathrm{ppm}$ relative to the zonal drag models with and without the $\sin\theta$ factor, respectively; see Figure \ref{fig:phasecurve}) with such a difference being small compared to changes that could result from other modelling choices (e.g., changes in metallicity, the inclusion of clouds).  There is also a reduction in the phase offset in the full drag phase curve relative to the zonal drag models due to the suppressed dynamics that results in the quasi-static region around the substellar point.

\subsection{The Magnetic Reynolds Number}

\begin{figure*}
    \centering
    \includegraphics{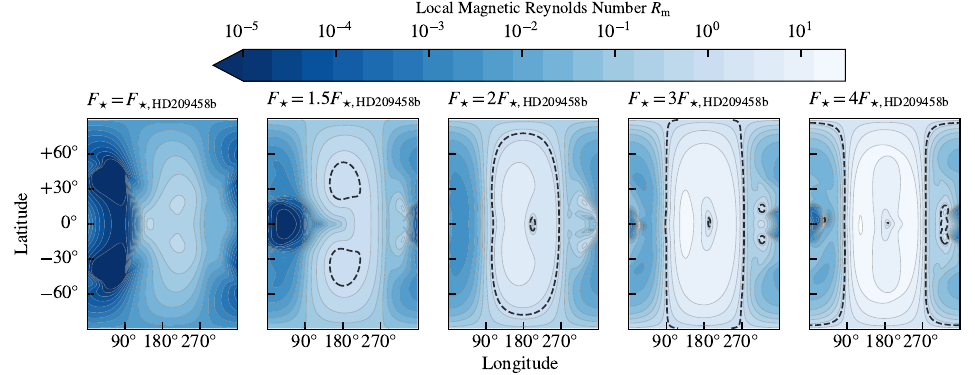}
    \caption{The magnetic Reynolds number $R_\mathrm{m}$ (Equation \ref{Eqn:Rm}) at a pressure of 1 mbar for each of the full drag simulations.  As the dayside temperatures increase with instellation the Reynolds number increases, hinting at the need to consider dayside advection of the magnetic field.}
    \label{fig:Rm}
\end{figure*}

The benefit of a magnetic drag model is that it can qualitatively capture the different coupling regimes between the dayside and nightside of a hot Jupiter.   That said, as discussed in Section \ref{Sec:Limitations}, as the dayside temperatures increase the underlying assumptions of the drag models breakdown.   Examining the magnetic Reynolds number (Equation \ref{Eqn:Rm}) in Figure \ref{fig:Rm}, the atmosphere remains in the $R_\mathrm{m}<1$ regime throughout for the fiducial case, while as the instellation increases an increasing fraction of the dayside  satisfies $R_\mathrm{m}>1$ while the nightside remains in the $R_\mathrm{m} < 1$ regime.   Furthermore, as discussed in Section \ref{Sec:Limitations}, the strong suppression of dayside motions within the drag model reduces $R_\mathrm{m}$, even as it remains greater than unity, further necessitating the proper consideration of the deformation of the magnetic field by the atmosphere.  

We thus stress that while higher instellation cases provide interesting insights into the effects of field geometry in the context of magnetic drag models and that they may capture some qualitative aspects of the day-night differences within the atmospheres, caution should be exercised in over-interpreting the results.

\subsection{Atmospheric Torque}

\begin{figure}
    \centering
    \includegraphics{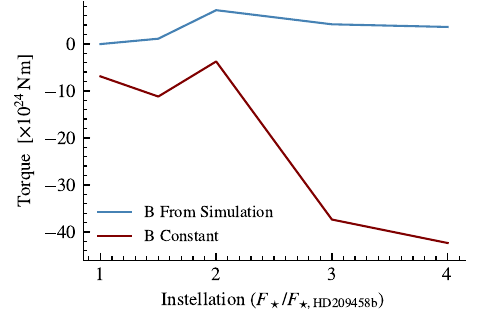}
    \caption{The torque on the atmosphere within the computational domain due to magnetic drag using the assumed magnetic field from the simulation  which varies throughout the computational domain (blue) and post-processing with an constant 10\,G magnetic field (red).}
    \label{fig:torque}
\end{figure}

Recent work by \citet{wazny_2025} has argued that the magnetic torque exerted on the atmosphere of a hot Jupiter can be sufficient to prevent it from reaching synchronous rotation, under the assumption that any torque exerted by the magnetic field on the upper atmosphere is offset by a commensurate torque on the unmodelled planetary interior.   We perform the same computation for the simulations found here.  

We compute the torque on the atmosphere within computational domain, $\Gamma_\mathrm{drag}$, by integrating over the localized magnetic torques over the domain $V$,

\begin{equation}
\Gamma_\mathrm{drag} = \int_V dV \rho \left(\frac{\partial u}{\partial t}\right)_\mathrm{magnetic} r\cos\theta = -\int_V dV \frac{\rho u}{\tau_\mathrm{drag}}r\cos\theta \,\, .
\label{Eqn:torque}
\end{equation}

\noindent {The torque for each simulation is shown in Figure \ref{fig:torque}, and we find for the suite of simulations presented here that the net torque on the computational domain is positive in all cases except the fiducial case where it is negative ($\Gamma_\mathrm{drag}=-5.26\times10^{22}\,\mathrm{N\, m}$) and two orders of magnitude smaller in magnitude than the torques in the more irradiates cases.   This differs from the results from \citet{wazny_2025} who found $\Gamma_\mathrm{drag}=-5.19\times 10^{24}\,\mathrm{N\,m}$ for their HD\,209458\,b simulation and negative torques for all other cases they investigated.  As the torque in our simulations is one percent of the value found in \citet{wazny_2025}, it is unlikely that it would result in meaningful departure from asynchronous rotation.  We also note that while the torque is negative at the instantaneous output shown, it oscillates between positive negative values of similar magnitude over the final 100 days, indicating that it may not exert a consistent negative torque, further reducing its impact. As the instellation increases, the net torque increases to consistently positive values $\sim 10^{24}\,\mathrm{N\, m}$, but not monotonically, likely due to formation of a quasistatic region around the substellar point which does not contribute significantly to the net torque.  If angular momentum is conserved between the atmosphere and the interior with no loss to space, as is assumed in \citet{wazny_2025}, these torques should result in a counter-rotating interior as opposed to the super-rotating interior.  This comes with a number of caveats.  Foremost, the magnetic field is unable evolve with the atmosphere, potentially to a state with a reduced net torque through the development of a toroidal component to the magnetic field leading to the torque on the atmosphere possibly being overestimated. Furthermore, the assumption that the angular momentum flux to space is zero is unmotivated, and it is unclear to what extent the torque on the atmosphere would result in a torque on the interior. }

While there are a number of differences between the models presented here and those of \citet{wazny_2025}, the likely reason for the differing torques is the assumed magnetic field strength. \citet{wazny_2025} assume a fixed value of the magnetic field strength ($B=10\,\mathrm{G}$) throughout the atmosphere, consistent with how zonal drag models have been used previously \citep[e.g.,][]{rauscher_2013,beltz_2022}, whereas here, despite the reference field strength being the same ($B_\mathrm{ref}=10\,\mathrm{G}$), the field strength varies throughout the computational domain (see Equation \ref{Eqn:B}) with stronger fields deep in the atmosphere and weaker fields near the outer boundary. When we instead post-process the results of our simulations assuming $B=10\,\mathrm{G}$ throughout the computational domain, with all other details remaining fixed, we find negative torques on the atmosphere within the computational domain for all our simulations and a torque of $\Gamma_\mathrm{red}=-6.9\times 10^{-24}\,\mathrm{N\,m}$ for the fiducial case, in particular, in qualitative agreement with \citet{wazny_2025}.  Moreover, post-processing with  field strengths of $2.5$ and $5\,\mathrm{G}$ similarly result in negative torques, indicating that the effect is, in part, geometric\footnote{In cases where the conductivity is Ohmic throughout the computational domain, the scaling of $\tau_\mathrm{drag}$ as $B^{-2}$ results in the torque in Equation \ref{Eqn:torque} scaling as $B^2$ and, as the dependence on the magnetic field can be moved outside of the integral, the sign of the torque remains the same for all values of $B$.  As the lowest pressures in the models presented here are not in the Ohmic limit, this may not hold more generally, but we note it nonetheless.}.  We caution that the torques computed through this post-processing are inconsistent with the drag forcing used in the simulations which may lead to conflicting results.   This is also true of the \citet{wazny_2025} calculation, where simulations from \citet{menou_2022} which do not include magnetic drag were used for their post-processing.   Without including the magnetic drag in the simulations, faster jets may result relative to a self-consistent simulation which then, during post-processing, cause an over-estimation of the torque on the jet.  This is especially true of situations where the flow is fundamentally changed by the inclusion of magnetic drag, such as in the highest instellation cases investigated here.  Although we find differing torques on the atmosphere, the underlying point of \citet{wazny_2025} that the net torque on the modelled atmospheres is non-zero remains.   

\section{Conclusions}
\label{Sec:Conclusions}

In this work, we have built upon prior works investigating zonal magnetic drag models \citep[e.g.,][]{rauscher_2013,beltz_2022} and have demonstrated that relaxing the implicit assumption of a zonal flow and the consideration of the three dimensional nature of the magnetic field geometry alters the flow from the more idealised models, except in the coolest, lowest ionisation cases, while the assumption of purely meridional currents consistently results in an underestimation in the magnetic drag, even in the lowest ionisation cases. We have only considered aligned dipoles within the analyses presented here, but the method  generalises to arbitrary field configurations and orientations. Deviations from zonal drag may still occur for misaligned dipoles or more complex geometries for the coolest of the hot Jupiters. As the temperatures increase in the upper atmosphere, a quasi-static region around the substellar point forms due to hot, ionized gas becoming trapped by the dipolar magnetic field.   The day side upwelling, normally peaking around the equator, instead forms an annulus at low pressures with the flow being directed towards the night side.  The atmosphere retains a deeper super-rotating jet; however, the jet is slowed to speeds $\lesssim 2\,\mathrm{km\,s^{-1}}$.

While the models presented here build upon prior work and relax some geometric assumptions, they are still subject to a number of limitations. The magnetic field is assumed to be a static dipole that did not evolve with the fluid. If this assumption were relaxed, the resulting magnetic field may develop a significantly different structure, potentially including a toroidal component as has been observed in non-deal magnetohydrodynamic simulations.  As such, while the approach presented in this study generalizes to arbitrary fixed magnetic field geometries, it remains simplified, and the results should still be viewed with a modicum of caution.  It remains, however, that current magnetic models of hot Jupiters -- both those that employ magnetic drag and those that solve the equations of non-ideal magnetohydrodynamics -- capture an incomplete picture of the dynamics in the presence of magnetic fields, and our understanding of what the true dynamics is requires results from our models to be framed in terms of the strengths and weaknesses of each approach.

\section*{Acknowledgements}
The authors would like to thank Cyril Gapp for valuable comments.  We would also like to thank the anonymous referee for their comments which we believe improved the paper greatly. DAC received funding from the Max Planck Society.  This research was also supported by a UK Research and Innovation (UKRI) Future Leaders Fellowship MR/T040866/1, and partly supported by the Leverhulme Trust through a research project grant RPG-2020-82 alongside a Science and Technology Facilities Council (STFC) Consolidated Grant ST/R000395/1.

The analysis of the simulation data made use of the following {\sc python} packages:  {\sc aeolus} \citep{sergeev_2022}, {\sc iris} \citep{hattersley_2023}, {\sc matplotlib} \citep{hunter_2007}, and {\sc numpy} \citep{harris_2020}.

\section*{Data Availability}

The simulation data are available for download from the Zenodo online repository at \href{https://doi.org/10.5281/zenodo.15832209}{doi.org/10.5281/zenodo.15832209}.  For the purpose of open access, the authors have applied a Creative Commons Attribution (CC BY) licence to any Author Accepted Manuscript version arising.



\bibliographystyle{mnras}
\bibliography{references} 




\appendix

\section{Ion-Neutral Drift Speeds in Hot Jupiter Atmospheres}
\label{Apdx:DriftSpeed}
As the partial coupling of the neutral gas to the magnetic field results in a relative motion between the ions and the neutrals, there is potential for observations of the relative velocities between ions and neutrals \citep[e.g.,][]{stangret_2022} to be used to infer the magnetic field strength, assuming the signals from the ions and the neutrals are both coming from the atmosphere \citep{savel_2024}.   The interpretation of these observations and the extraction of a magnetic field strength requires an accurate underlying model of how the ions should move relative to the neutrals in the exoplanet atmosphere.  In this appendix, we derive the speed of ions relative to the neutrals in the resistive, single-fluid limit. 

For the weakly ionized atmospheres considered here, the ions are expected to contribute negligibly to the total momentum and are assumed to be in a force balance between the Lorentz forces and the collisional drag forces between the ions and the neutrals\footnote{The assumption of negligible momentum in the ions can be relaxed at the expense of algebraic complexity (see, e.g., \citealt{koskinen_2014}).  Small scale waves will invariably decouple from the neutrals, but these are not considered here.},

\begin{equation}
0 = q_s n_s\left(\mathbf{E_n} + \frac{\mathbf{w}_s}{c}\times \mathbf{B}\right) + \frac{n_sm_s}{\tau_{s\mathrm{n}}}\mathbf{w}_s\,\, ,
\label{Eqn:force_balance}
\end{equation}

\noindent where $\mathbf{w}_s$ is the  velocity of ion species $s$
relative to the neutral gas.  With a bit of algebraic manipulation of Equation \ref{Eqn:force_balance}\footnote{Applying the cross-product with $\mathbf{b}$ once and twice to Equation \ref{Eqn:force_balance} yields two equations that can be used to solve for $\mathbf{w}_{s,\perp}$.}, the perpendicular component of the drift can be shown to be 

\begin{equation}
\mathbf{w}_{s,\perp} = \frac{q_s\tau_{s\mathrm{n}}}{m_s}\left(1 + \omega_s^2\tau_{s\mathrm{n}}^2 \right)^{-1}\left(-\mathbf{E}_{n,\perp} + \omega_s\tau_{s\mathrm{n}}\mathbf{E}_\mathrm{n}\times \mathbf{b}\right)\,\,.
\label{Eqn:wvsE}
\end{equation}

\noindent The magnetization of species $s$ is $M_s = \omega_s \tau_{s\mathrm{n}}$ and serves as an indicator of the relative coupling of the species to the magnetic field or the neutral gas.   Noting that $\mathbf{E}_{n,\perp}$ and $\mathbf{E}_\mathrm{n}\times \mathbf{b}$ are perpendicular and making use of Ohm's law (Equation \ref{Eqn:OhmsLaw}), the magnitude of the drift can be written as

\begin{align}
|w_{s,\perp}| & = \frac{|q_s|\tau_{s\mathrm{n}}}{m_s}\left(1 + \omega_s^2\tau_{s\mathrm{n}}^2 \right)^{-1/2}|E_{\mathrm{n},\perp}| \\
& =  \frac{|q_s|\tau_{s\mathrm{n}}}{m_s\sqrt{(1 + \omega_s^2\tau_{s\mathrm{n}}^2)(\sigma_\perp^2 + \sigma_\mathrm{H}^2)}}|j_\perp|\,\, .
\label{Eqn:wvsEmag}
\end{align}

\noindent Using Equations \ref{Eqn:etaparl}-\ref{Eqn:etaH} and \ref{Eqn:Induction}, the magnitude of the perpendicular current can be estimated to be $|j_\perp|\simeq \sqrt{\sigma_\perp^2 + \sigma_\mathrm{H}^2} B u_\mathrm{n}$ where $u_\mathrm{n}$ is the speed of the neutral gas relative to the magnetic field, which is generalised from the assumption in the Ohmic limit that $|j| = \sigma B u_\mathrm{n}$ used in \citet{perna_2010a}, \citet{koskinen_2014},  and \citet{savel_2024}. This reduces to

\begin{equation}
|w_{s,\perp}| = \frac{|\omega_s|\tau_{s\mathrm{n}}}{\sqrt{1+\omega_s^2\tau_{s\mathrm{n}}^2}} u_\mathrm{n}\,\, .
\label{Eqn:wperpmag}
\end{equation}

\noindent It can be seen immediately, as a consequence of $|\omega_s|\tau_{s\mathrm{n}} > 0$, that the drift does not exceed the neutral velocity, with $|w_{s,\perp}| = 0$ corresponding to ion species $s$ moving with the neutrals and $|w_{s,\perp}| = u_\mathrm{n}$ corresponding to ion species $s$ remaining attached to the magnetic field.   This can be instead be expressed in a vector form via the introduction of the drift of the field relative to the neutrals, $\mathbf{w}_\mathrm{f}$\footnote{In the case where the magnetic field is stationary in the frame of the planet, as is assumed in the magnetic drag models discussed in the main text, this is equivalent to $\mathbf{w}_\mathrm{f,\perp} = -\mathbf{u}_\perp$.} following the derivation from \citet[][Appendix C]{kunz_2009}, resulting in the velocity of species $s$ relative to the neutrals being

\begin{equation}
\mathbf{w}_{s,\perp} = \frac{\omega_s^2\tau_{s\mathrm{n}}^2}{1+\omega_s^2\tau_{s\mathrm{n}}^2} \mathbf{w}_\mathrm{f,\perp} - \frac{\omega_s\tau_{s\mathrm{n}}}{1+\omega_s^2\tau_{s\mathrm{n}}^2} \mathbf{w}_\mathrm{f}\times\mathbf{b},
\label{Eqn:wperpvect}
\end{equation}

\noindent which satisfies the condition in Equation \ref{Eqn:wperpmag}.  

This differs from the conclusions of \citet{savel_2024} where the ion drift speed is unbounded.  The discrepancy arises from the assumption in \citet{savel_2024} that 

\begin{equation}
\frac{1}{c}\mathbf{j}\times \mathbf{B} \approx -\frac{n_s m_s}{\tau_{s\mathrm{n}}}\mathbf{w}_s\,\, ,
\label{Eqn:jxbsavel}
\end{equation}

\noindent holds for all species $s$ (their Equation 1).  It can be arrived at by summing Equation \ref{Eqn:force_balance} over all ionised species and imposing charge neutrality and substituting for the current $\mathbf{j}=\sum_s n_sq_s \mathbf{w}_s$, yielding 

\begin{equation}
\frac{1}{c}\mathbf{j}\times \mathbf{B} = -\sum_s\frac{n_s m_s}{\tau_{s\mathrm{n}}}\mathbf{w}_s\,\,.
\label{Eqn:jxbsum}
\end{equation}

\noindent While it may hold that a single species dominates the sum in Equation \ref{Eqn:jxbsum} thus allowing for Equation \ref{Eqn:jxbsavel} to hold true, it is not required in general and does not hold true for all species, and thus cannot be used for trace species such as barium ions.   


\bsp	
\label{lastpage}
\end{document}